\documentclass[10pt]{paper} 
\usepackage[english]{babel}

\usepackage{a4wide}
\usepackage[a4paper,nofoot,body={16cm,23cm}]{geometry}
\usepackage{amsmath, amsfonts, amsthm, amssymb,verbatim} 
\usepackage{graphicx}
\usepackage{xspace,color} 
\usepackage{enumitem} 
\usepackage{xspace} 
\usepackage{cite}
\newtheorem{theorem}{Theorem}[section]
\newtheorem*{theorem*}{Theorem}

\numberwithin{equation}{section}

\usepackage[bookmarks=true,
           bookmarksopen=true,hidelinks]{hyperref}

\newcommand \bel {\begin{equation}\label}
\newcommand \bei {\begin{itemize}}
\newcommand \eei {\end{itemize}}

\newcommand \bse {\begin{subequations}}
\newcommand \ese {\end{subequations}}
 
\newcommand \bz {\begin{itemize}}
\newcommand \ez {\end{itemize}}
\newcommand \ben {\begin{enumerate}}
\newcommand\een {\end{enumerate}} 
\newcommand \R {\mathbb R}
\newcommand{\Eqref}[1]{Eq.~\eqref{#1}}
\newcommand{\Eqsref}[1]{Eqs.~\eqref{#1}}
\newcommand{\Sectionref}[1]{Section~\ref{#1}}  
\newcommand{\Sectionsref}[1]{Sections~\ref{#1}}

\newcommand{\Theoremref}[1]{Theorem~\ref{#1}}

\newcommand{\Figref}[1]{Figure~\ref{#1}}
\newcommand{\keyword}[1]{\textit{#1}}

\newcommand \muh M 
\newcommand \vv V

\newcommand \ub {\overline u}

\newcommand \del 		\partial
\newcommand \eps 		\epsilon
\newcommand \lam 		\lambda 
\newcommand \be 		{\begin{equation}}
\newcommand\ee 		{\end{equation}}

\newcommand \tildeT {\widetilde T}

\newcounter{mnotecount}[section]
\let\oldmarginpar\marginpar
\renewcommand\marginpar[1]{\-\oldmarginpar[\raggedleft\footnotesize #1]%
 {\raggedright\footnotesize #1}}

\graphicspath{{.}{Figures/}}

\newcommand\blfootnote[1]{%
  \begingroup
  \renewcommand\thefootnote{}\footnote{#1}%
  \addtocounter{footnote}{-1}%
  \endgroup
}

\newcommand{\Vb}{\boldsymbol V}
\newcommand{\VbLOT}{\boldsymbol V_*}
\renewcommand{\ub}{\boldsymbol u}
\newcommand{\fb}{\boldsymbol f}
\newcommand{\yb}{\boldsymbol y}
\newcommand{\VbCD}{\boldsymbol V_0}
\newcommand{\Gb}{\boldsymbol G}

\begin{document}

\title{A numerical algorithm for Fuchsian equations and fluid flows on cosmological spacetimes} 
 
\author{Florian Beyer$^1$ and Philippe G. LeFloch$^2$}

\date{}

\maketitle

\blfootnote{$^1$ Department of Mathematics and Statistics, University of Otago, P.O.~Box 56, 
 Dunedin 9054, New Zealand. Email: {\tt fbeyer@maths.otago.ac.nz}
 \newline 
 $^2$ Laboratoire Jacques-Louis Lions and Centre National de la Recherche Scientifique, 
 Sorbonne Universit\'e, 4 Place Jussieu, 75252 Paris, France. Email: {\tt contact@philippelefloch.org} 
\hfill
Completed in May 2020. 
}

\begin{abstract} We consider a class of Fuchsian equations that, for instance, describes the evolution of compressible fluid flows on a cosmological spacetime. Using the method of lines, we introduce a numerical algorithm for the singular initial value problem when data are imposed on the cosmological singularity and the evolution is performed {\sl from} the singularity hypersurface. 
We approximate the singular Cauchy problem of Fuchsian type by a sequence of regular Cauchy problems, which we next discretize by pseudo-spectral and Runge-Kutta techniques. 
Our main contribution is a detailed analysis of the numerical error which has two distinct sources, and our main proposal here is to {\sl keep in balance} the errors arising at the continuum and at the discrete levels of approximation. We present numerical experiments which strongly support our theoretical conclusions. This strategy is finally applied to applied to compressible fluid flows evolving on a Kasner spacetime, and we numerically demonstrate the nonlinear stability of such flows, at least in the so-called sub-critical regime identified earlier by the authors. 
\end{abstract}

\setcounter{tocdepth}{1} 
\tableofcontents


\section{Introduction}
\label{sec:intro}

\paragraph{Cosmological singularities.} 

We introduce here a numerical algorithm for computing and investigating a problem about the relativistic Euler equations of compressible fluids, which arises in cosmology. 
Our conclusions below should be relevant also for other problems involving Fuchsian-type partial differential equations.  
In cosmology, two distinguished times can be used for describing the history of the universe, that is, the time at which we make our measurements and observations, or the time of the ``big bang'' at which the history of the universe started. 
Numerical calculations for cosmological models can thus be performed by choosing either of these two times as 
our \emph{initial time} at which suitable initial conditions are prescribed. In the first case, this leads us to the (regular) Cauchy problem for the Einstein-Euler equations and is the standard formulation for wave-type partial differential equations (PDEs). 
In the second case, this leads us to the so-called \emph{singular Cauchy problem} for which the initial conditions are prescribed at the singularity initial time (denoted below by $t=0$) and whose solution are initially singular -- in stark contrast to the regular Cauchy problem.

\paragraph{Properties of singular solutions} 

The theory of the regular Cauchy problem is far more developed than that of the singular Cauchy problem. We focus on the latter and investigate numerical issues that arise with Fuchsian equations. We do not attempt here to review the mathematical theory, and will only review below the material that will be specifically needed for our study.  
Theoretical advances can be found in \cite{ames2013a,andersson2001,beyer2010b,beyer2012a,beyer-PLF-2017,choquet-bruhat2004,rendall2000}, including the derivation of expansions for solutions in the vicinity of a cosmological singularity.
Of course, in most cases the solutions are not known in a close form (with the exception of, for instance, \cite{beyer2014,hennig2016,hennig2016a,hennig2019}) and therefore we must appeal to numerical approximation in order to analyze the qualitative properties of solutions. Since the solutions of interest blow-up on the singularity $t=0$, it it necessary to develop an adapted methodology in order to reach reliable conclusions on, for instance, the stability of solutions with respect to their initial data. 
 
\paragraph{Numerical strategy} 

Motivated by the work \cite{amorim2009} on the so-called Gowdy equations of general relativity (describing the spacetime geometry in presence of two commuting killing fields), we initiated in \cite{beyer2010b} the development of numerical schemes 
for singular PDEs of Fuchsian and wave-like type, and successfully applied theem to the Einstein equations; 
an application was next presented in \cite{beyer2011}. The basic idea in \cite{beyer2010b} is to approximate the singular initial value problem by a {\sl sequence of regular Cauchy problems,} each of them being next approximated numerically. 
We will revisit this strategy in  \Sectionref{sec:errorestimates} below; interestingly, this essentially mimics an abstract technique used first for establishing an existence theory for Fuchsian equations. 

In order to control the convergence of such an approximation scheme we need to control two distinct sources of error.  
\begin{enumerate}

\item The \emph{continuum approximation error} is produced by approximating the {singular initial value problem} by a sequence of regular Cauchy problems. 

\item The \emph{numerical approximation error} is produced by numerically approximating the solution of each Cauchy problem under consideration. 
\end{enumerate}
The sum of these errors will be referred to here as the \emph{total approximation error}, and our main purpose is to estimate this error, and then apply our conclusions to fluid flows in the vicinity of a cosmological singularity. 

For the applications considered in \cite{beyer2010b,beyer2011} it was sufficient to work under the assumption that the numerical approximation error is negligible in comparison to the continuum approximation error. Namely, this is reasonable in the regime where the numerical resolution so high that numerical solutions can essentially be treated as exact solutions. 
However, for more complex applications (such as the nonlinear stability with respect to singularity initial data, treated in Section 6 below)
this assumption (of sufficiently high numerical resolution) is prohibitive, and this is the starting point of our analysis. 

In order to address this issue,  we develop a systematic treatment of the error sources. We rely on the {\sl method of lines} (see \cite{boyd1989,leveque1990}) and approximate the solution to the Euler equations by a (large) system of ordinary differential equations (cf. \Sectionref{sec:mainnumapplications}) which, on a cosmological background spacetime, is precisely of Fuchsian-type. 

\paragraph{Main purpose in this paper}

After stating the problem in Section 2, in \Sectionref{sec:FuchsianTheory} we start by discussing the relevant class of singular ordinary differential equations which is the main focus of the present paper.
A central idea behind our approach is the following one. Instead of assuming that the numerical approximation error is negligible,
 we study this error and eventually discover that the optimal strategy is to \emph{balance the numerical and continuum approximation errors},  i.e.~to make both errors roughly of the same order in magnitude during the whole evolution. Standard adaptive numerical ODE evolution schemes (such as in \cite{gautschi2012}) are not directly suitable for our purpose, 
 since they are designed to achieve a different accuracy goal. In our setup in order to achieve the desired balance, 
 we must take the theoretical decay (or blow up) rates of the solutions into account and, indeed, we derive estimates for the numerical approximation error which are linked to the fact that the solution is \emph{singular} at $t=0$.
 
  As a case study, we choose to analyze the second-order Runge-Kutta (RK2) scheme, but we expect similar arguments to apply to other classes of discrete schemes. We formulate our main results in \Theoremref{theorem:RK2} below. 
  In a second stage of the analysis, we propose to balance these estimates (for the numerical approximation error) 
  with (more direct) estimates that we derive for the continuum approximation error; 
  see \Theoremref{thm:approx} below. 

Let us point out that all of the errors terms are measured with respect to a {\sl one-parameter family of weighted norms},
 involving a weight $t^\lambda$, where $\lambda$ is a problem-dependent exponent. 
The larger $\lambda$ is, the more this weight penalizes slow decay (or even blow-up) in the limit $t \searrow 0$. The freedom of choosing $\lambda$ appropriately allows us to control the decay (or blow up) rates of the individual error components relative to the theoretically known decay (or blow up) rate of the actual solution at $t=0$. We refer to \Sectionsref{sec:onestepmethods} and \ref{sec:mainanalyticalresult} for further details.

\paragraph{Asymptotically balanced discretizations}

In order to ensure a suitable balance property, we impose a relationship between the time step size $h$ (of the numerical approximation) and the initial time $t_*$ arising in approximating the singular Cauchy problem. (Note that non-constant time step sizes can be treated, as discussed in \Sectionref{sec:mainanalyticalresult}.)  
This relationship is of the form 
\[
h\propto t_*^{1+\beta},
\] 
in which $\beta$, in principle, is an arbitrary constant. 
While our rigorous analysis is restricted to the case $\beta\ge 0$, our numerical experiments (in \Sectionref{sec:Numtestproblem}) suggest that picking up $\beta<0$ is also possible, but does not yield any practical advantage. 

In the case $\beta\ge0$ and for RK2 at least, in \Sectionsref{sec:mainanalyticalresult} and \ref{sec:proof}  we establish that the numerical and the continuum approximation errors are {\bf asymptotically balanced provided} $\beta$ is chosen to be 
\[
\beta_* = (\delta-\lambda)/2-1,\] 
(or, if this is negative, $\beta_*=0$), in which the parameter 
$\delta$ is the ``theoretical decay exponent'' determined by the singular initial value problem at $t=0$; see Section \ref{thm:SIVP}.

The size of the parameter $\lambda$ has a significant effect here. We also find that if we pick $\beta$ smaller than $\beta_*$
 the total approximation error is {\sl asymptotically dominated} by the numerical one and the solution is therefore not resolved. 
 If $\beta$ is picked to be larger than $\beta_*$, 
 the total approximation error turns out to be {\sl asymptotically dominated by the continuum error} and 
 some of the numerical effort is ``wasted''.  
 Choosing $\beta$ too large is never beneficial. 
 The possibility of choosing $\beta$ smaller than the balanced value $\beta_*$ can still be exploited in practice, 
 since it allows to improve the computation work at the price of reducing the numerical accuracy. 
 In fact, this cost and this benefit can cancel each other, and this justifies that we may sometime work with the choice $\beta=0$. 

\paragraph{The role of Fuchsian transformations}

Certain classes of transformations leave invariant the Fuchsian form of the singular equations, as is discussed in \Sectionref{sec:properties} below. Despite the numerical schemes under consideration may
 be not invariant under such transformations, we nevertheless show 
 that our notions of asymptotic balance and asymptotic efficiency above \emph{are invariant.} 
Numerical approximation schemes can therefore not simply be made ``higher order'' by applying such a transformation. 
In fact, one approach to designing higher-order schemes for the singular problem 
under consideration consists of transforming the singular initial value problem into a more regular one by  subtracting an expansion of the solution. Such a theory of \emph{higher order expansions} was discussed earlier in \cite{beyer2010b, beyer2011} as well as \cite{ames2013a}.

The conclusions reached in \Sectionref{sec:errorestimates} are investigated numerically in \Sectionref{sec:Numtestproblem} and finally allow us to address an issue concerning the Euler equations on a cosmological background near the big bang singularity; see 
 \Sectionref{sec:mainnumapplications}. The behavior of such a fluid flow was discussed theoretically in \cite{beyer-PLF-2017,beyer2019a}, but open questions remained concerning the asymptotic stability. Our numerical results provide strong support that the singular behavior of the fluids at the cosmological singularity is dynamically stable.


\section{Compressible fluid flows on a Kasner background}

\subsection{ Formulation of the problem}

\paragraph{Euler equations}

In this section we consider the Euler equations using the formalism in
\cite{frauendiener2003,walton2005}; see also \cite{beyer-PLF-2017}. 
Perfect fluids can be represented by a (in general
not normalized) 4-vector field $V$ satisfying the following quasilinear  symmetric hyperbolic system
\begin{equation}
\label{eq:AAA1}
0 ={A^\delta}_{\alpha\beta}\nabla_\delta V^\beta,
\quad 
A^\delta_{\alpha\beta}
 =\frac{3\gamma-2}{\gamma-1} \frac{V_\alpha V_\beta}{V^2} V^\delta+V^\delta g_{\alpha\beta}
  +2{g^\delta}_{(\beta} V_{\alpha)},\quad
V^2=-V_\alpha V^\alpha.
\end{equation}
Here we use the Einstein summation convention over repeated indices. Indices are lowered and
raised with the (so far arbitrary) Lorentzian spacetime metric $g$.
This system is equivalent to the Euler equations for perfect fluids if we impose the linear equation of state
$
  P=(\gamma-1)\rho,
$
where $P$ is the fluid pressure, $\rho$ is the fluid density. Here, the
\emph{speed of sound} $c_s$  is a constant, namely 
\begin{equation}
  \label{eq:speedsoundrestr}
  c_s^2=\gamma-1\in (0,1).
\end{equation}
The normalized fluid 4-vector field $U$, the fluid pressure $P$ and
the fluid density $\rho$ are recovered from the $4$-vector field $V$ as follows:
\begin{equation}
  \label{eq:physicsquantitiesfluid}
U=\frac{V}{\sqrt{V^2}},\qquad
P= (V^2)^{-\frac {\gamma}{2(\gamma-1)}},
\qquad \rho=\frac{1}{\gamma-1} (V^2)^{-\frac {\gamma}{2(\gamma-1)}}.
\end{equation}

\paragraph{Spacetime geometry of interest}

The fluid is assumed to evolve on a \emph{Kasner spacetime}, which is a spatially homogeneous (but possibly 
 anisotropic) solution $(M,g)$ to Einstein's vacuum equations with\footnote{Observe that the time variables $t$ differs from the one in
  \cite{beyer2019a} by a minus sign. Here we always assume that
  $t>0$ as for example in \cite{beyer-PLF-2017}.}  $M=(0,\infty)\times\Sigma$ with $\Sigma=\mathbb T^3$ and
\begin{equation}
\label{Kasner-k}
g = t^{\frac{K^2-1}{2}} \big( - d t\otimes dt + dx\otimes dx \big) + t^{1-K} dy\otimes dy +t^{1+K} dz\otimes dz. 
\end{equation}
Here, we take $t \in (0,\infty)$ and $x,y,z \in (0,2\pi)$. The free parameter $K\in\mathbb R$  is often referred to as the \emph{asymptotic velocity}. 
Except for the three flat Kasner cases given by
$K=1$, $K=-1$, and (formally) $|K|\to  \infty$, the Kasner
metrics $g$ have a curvature singularity in the limit $t\searrow 0$. 

Observe that the coordinate transformation 
 \[\widetilde t=\frac{4}{K^2+3} (-t)^{\frac{K^2+3}4},\quad 
 \widetilde x=\left(\frac{K^2+3}{4}\right)^{\frac{K^2-1}{K^2+3}} x,\quad
 \widetilde y=\left(\frac{K^2+3}{4}\right)^{\frac{2(1-K)}{K^2+3}} y,\quad
 \widetilde z=\left(\frac{K^2+3}{4}\right)^{\frac{2(1+K)}{K^2+3}} z,\]
 brings this metric to the more conventional form 
\be
 g = -d\widetilde t\otimes d\widetilde t + \widetilde t^{2p_1} d\widetilde x\otimes
 d\widetilde x  + \widetilde t^{2p_2} d\widetilde y\otimes d\widetilde y  + \widetilde t^{2p_3} d\widetilde z\otimes d \widetilde z,
 \ee
 where 
\be
  p_1 =(K^2-1)/(K^2+3), \quad
  p_2 =2(1-K)/(K^2+3),\quad
  p_3 =2(1+K)/(K^2+3),
\ee
 are the \emph{Kasner exponents}.
 These satisfy the \emph{Kasner relations} $\sum_i p_i=0$ and $\sum_i p^2_i=1$.

\paragraph{The first-order formulation}

For simplicity, we thus
focus on the Euler equations
on a fixed Kasner spacetime, but most of the ideas in the present paper should carry over to the \emph{coupled}
Einstein-Euler system (considered for example in \cite{beyer-PLF-2017}).
Restricting to the same symmetry class considered in \cite{beyer-PLF-2017}, we assume, with respect to the coordinates $(t,x,y,z)$ on the background Kasner spacetime $(M,g)$, that (I) the vector fields $\partial_y$ and $\partial_z$ are symmetries of the fluid, i.e., $[\partial_{y},V]=[\partial_z,V]=0$, and that (II) the fluid only flows into the $x$-direction\footnote{It was observed in \cite{lefloch2011} that (II) necessarily follows from (I) in the case of the \emph{coupled} Einstein-Euler system. Since the background spacetime is fixed here, however, we could consider only assumption (I), but not (II).} $dy(V)=dz(V)=0$. The fluid variables of interest are therefore the two non-trivial coordinate components of the vector field $V=(V^0(t,x), V^1(t,x),0,0)$.
Under these assumptions, the Euler equations \eqref{eq:AAA1} take the form
\bse
  \begin{equation}
    \label{eq:Eulereqssymmhyp1}
    \begin{split}
    \bar B^0(V^0,V^1) &\partial_t \begin{pmatrix}
      V^0\\
      V^1
    \end{pmatrix}
    +\bar B^1(V^0,V^1) \partial_x \begin{pmatrix}
      V^0\\
      V^1
    \end{pmatrix}=
    \bar G(t,V^0,V^1)
  \end{split}
\end{equation}
with
\begin{align}
  \label{eq:Eulereqssymmhyp1B0}
  \bar B^0(v^0,v^1)
  &=\begin{pmatrix}
  {v^0} \left((v^0)^2+3 (v^1)^2 (\gamma -1)\right) & {v^1}
   \left((v^0)^2 (1-2 \gamma )-(v^1)^2 (\gamma -1)\right) \\
 {v^1} \left((v^0)^2 (1-2 \gamma )-(v^1)^2 (\gamma -1)\right) &
   {v^0} \left((\gamma -1) (v^0)^2+(v^1)^2 (2 \gamma -1)\right)
 \end{pmatrix},\\
  \label{eq:Eulereqssymmhyp1B1}
  \bar B^1(v^0,v^1)
  &=-\begin{pmatrix}
    -{v^1} \left((1-2 \gamma) (v^0)^2-(v^1)^2 (\gamma -1)\right) &
   -{v^0} \left((v^0)^2 (\gamma -1) -(v^1)^2 (1-2 \gamma )\right) \\
 -{v^0} \left((v^0)^2 (\gamma -1) -(v^1)^2 (1-2 \gamma )\right) &
   {v^1} \left(3 (\gamma -1) (v^0)^2+(v^1)^2\right)
  \end{pmatrix},\\
  \label{eq:Eulereqssymmhyp1G}
  \bar G(t,v^0,v^1)&=\frac{\Gamma}t
  ((v^0)^2 - (v^1)^2)\begin{pmatrix}
    (v^0)^2 \\
    v^0 v^1 
  \end{pmatrix},
\end{align}
\ese
where the constant $\Gamma$ is defined as
\begin{equation}
  \label{eq:defGamma}
  \Gamma=\frac 14\left(3 \gamma-2 - K^2 (2 - \gamma)\right).
\end{equation}
It is useful to observe that
\begin{equation}
  \label{eq:Gammabound}
  \Gamma<1
\end{equation}
follows from \eqref{eq:speedsoundrestr}. 
Motivated by the evidence proposed in  \cite{beyer-PLF-2017}, we expect that the fluid flow is in general \emph{dynamically unstable} if $\Gamma\le 0$, so that we naturally restrict attention to all $\gamma\in (1,2)$ and $K\in\mathbb R$ such that 
\begin{equation}
  \label{eq:Gammabound2}
  \Gamma>0.
\end{equation}
In the terminology of \cite{beyer-PLF-2017}, this is the \emph{sub-critical} case, as opposed to the (super)-critical cases $\Gamma\le 0$.


\subsection{The Fuchsian structure}

\paragraph{Expansion on the cosmological singularity}

The rigorous analysis in \cite{beyer-PLF-2017} for the
\emph{coupled} Einstein-Euler case also applies to the Euler equations on a \emph{fixed} Kasner
background. In fact, the results obtained about the asymptotics in are slightly stronger in the present context. One can show that for each (smooth) positive function $V^0_*(x)$ and function $V^1_{*}(x)$,
there exists a time $T>0$ and a (smooth) solution
$V_{\mathrm{SIVP}}=(V_{\mathrm{SIVP}}^0(t,x),
V_{\mathrm{SIVP}}^1(t,x))$ of
\Eqsref{eq:Eulereqssymmhyp1}--\eqref{eq:Eulereqssymmhyp1G} defined on
$(0,T]\times\mathbb T^1$ such that
\begin{equation}
  \label{eq:fluidspecialasympt3}
  V_{\mathrm{SIVP}}^0(t,x)=V^0_*(x) t^\Gamma+u^0(t,x),\quad
  V_{\mathrm{SIVP}}^1(t,x)=V^1_{*}(x) t^{2\Gamma}+u^1(t,x). 
\end{equation}
Here, $u=(u^0,u^1)$ are uniquely determined by the
condition that $\sup_{t\in(0,T]}\|t^{-\mu_0} u(t,\cdot)\| < + \infty$ for
any $\mu_0\in (2\Gamma,3\Gamma)$. A particular example of such a solution is given
by $V^0_*=\mathrm{const}>0$ and $V_{*}^1=0$ in which case $u^0=u^1=0$. In
consistency with \Sectionref{sec:FuchsianTheory}, we interpret
$V_*=(V_*^0,V_*^1)$ as \emph{asymptotic data} and $u$ as the unknown of
the singular initial value problem. Since the
Euler equations are a system of PDEs, the theory for ODEs in \Sectionref{sec:FuchsianTheory} below should be applied after a spatial discretization in space is also performed. This issue is discussed in \Sectionref{sec:mainnumericalsetup} below. 

\paragraph{Nonlinear stability}

The main question we tackle here is whether a sub-critical
family of solutions to the Euler equations \eqref{eq:Eulereqssymmhyp1}--\eqref{eq:Eulereqssymmhyp1G} enjoying with asymptotics
\Eqref{eq:fluidspecialasympt3} is asymptotically stable, in a sense that will be made clear below.  
In fact, the recent results in
\cite{beyer2019a} suggest asymptotic stability in a rather strong sense.  Without going into
the technical details now, the stability result can be summarized as follows.

Fix $\gamma$ and $K$ as above so that $\Gamma>0$.
Pick arbitrary smooth asymptotic data $V_*=(V_*^1,V_*^2)$ with $V_*^1>0$ and let $V_{\mathrm{SIVP}}$ be  the uniquely determined
solution of the Euler equations \eqref{eq:Eulereqssymmhyp1}--\eqref{eq:Eulereqssymmhyp1G}
with the asymptotics in \Eqref{eq:fluidspecialasympt3} defined on the time
interval $(0,T]$. Then pick arbitrary smooth Cauchy data $V_0$,
 and let $V$ be the unique solution of the Cauchy problem of the Euler
 equations \eqref{eq:Eulereqssymmhyp1}--\eqref{eq:Eulereqssymmhyp1G} for the same choice of $\gamma$ and $K$ with Cauchy data $V_0$ imposed at $t=T$. It was shown in \cite{beyer2019a} that $V$ can be extended
 all the way down to $t=0$ (global existence) provided $V_0$
is sufficiently close (in a certain sense, see below) to $V_{\mathrm{SIVP}}(T,\cdot)$,
and, the asymptotics of $V$ at $t=0$ is
\emph{similar} to that of $V_{\mathrm{SIVP}}$ in the sense that the limit
\be
\lim_{t\searrow 0} t^{-\Gamma}V^0(t,\cdot)=V^0_\infty
\ee 
exists (in analogy to the first relation in \Eqref{eq:fluidspecialasympt3}) and
the size of $V^0_*-V^0_\infty$ 
is bounded by the size of $V_0-V_{\mathrm{SIVP}}(T,\cdot)$ (in a specific sense which we do not discuss here). This
therefore yields a notion of asymptotic stability of 
that sub-critical
family of solutions of the Euler equations \eqref{eq:Eulereqssymmhyp1}--\eqref{eq:Eulereqssymmhyp1G} with the asymptotics
\Eqref{eq:fluidspecialasympt3} constructed in \cite{beyer-PLF-2017}  under sufficiently small nonlinear perturbations.
Observe however that the result in \cite{beyer2019a} is not strong enough to
conclude anything about the limit of $t^{-2\Gamma}V^1(t,\cdot)$, i.e., it does not provide an analogue of the \emph{second} relation in \Eqref{eq:fluidspecialasympt3}. The theoretical results of asymptotic stability therefore \emph{do not characterize the asymptotics} of $V^1$. 

\paragraph{An open problem}

One aim of the present paper is precisely to \emph{rely on numerical investigations and
  provide strong evidence} that a sub-critical
family of solutions of the Euler equations \eqref{eq:Eulereqssymmhyp1}--\eqref{eq:Eulereqssymmhyp1G} with the asymptotics
\Eqref{eq:fluidspecialasympt3} constructed in \cite{beyer-PLF-2017}  satisfies a notion of asymptotic stability concerning \emph{both components $V^0$ and $V^1$}. 
In particular we want to demonstrate that the limit 
\be
\lim_{t\searrow 0} t^{-2\Gamma}V^1(t,\cdot)=V^1_\infty
\ee
  exists (in the regime of sufficiently small perturbations) and that $V^1_{*}-V^1_\infty$ is controlled by the
 $V_0-V_{\mathrm{SIVP}}(T,\cdot)$.


\section{The singular initial value problem for Fuchsian equations}
\label{sec:FuchsianTheory}

\subsection{Basic model of interest}
\label{thm:SIVP}

\paragraph{Fuchsian equations}

We begin our analysis with the following class of ordinary differential equations (ODEs) 
\begin{equation}
  \label{eq:FuchsianODEgen}
  t\partial_t u(t)-A u(t) = f(t,u(t)),\quad t>0,
\end{equation}
where  $u=u(t)$ is an $n$-vector-valued unknown while a constant $n\times n$-matrix $A$ and an $n$-vector valued function $f=f(t,y)$ (enjoying certain regularity properties, specified below) are prescribed. 
Such equations having a $1/t$-singularity at $t=0$ (after the equation is
divided by $t$) is called a \keyword{Fuchsian ODE}.  We are interested in the so-called \keyword{singular initial
  value problem} for \Eqref{eq:FuchsianODEgen}, that is, we solve this equation forward in time by evolving {\sl from} the singularity point $t=0$ with suitable \emph{asymptotic data} prescribed at $t=0$. For this problem, the existence and uniqueness of a solution is standard and we summarize the results here first. In what follows, $|\cdot|$ denotes the Euclidean norm for vectors
in $\R^n$.

\paragraph{The singular initial value problem}
    
Our setup is as follows. 
    Pick a time $T\in (0,1]$, a constant $s>0$ and an $n\times n$-matrix $A$, and let $\mu_0\in\R$ be such that\footnote{$I$ denotes the $n\times n$-identity matrix.} $\mu_0 I - A$ is positive definite, and consider also a smooth
    function
$f: D\to \R^n$ defined in an open subset $D$ of $\R^{n+1}$, 
  such that
\[
  \big\{ (t,y)\in (0,T]\times\R^n\,  / \, |y|\le s \, t^{\mu_{0}} \big\}\subset D.
\]
This function $f$ (arising in the right-hand side of \eqref{eq:FuchsianODEgen}) is assumed to have the following behavior near the singularity for some exponent $\delta>\mu_0$:
  \begin{enumerate}[label=(\roman{*}),leftmargin=*]
  
  \item Given any function $y:(0,T]\to \R^n$ with $\sup_{t\in(0,T]}|t^{-\mu_0} y(t)|\le s$, then
    \begin{equation}
      \label{eq:BoundSouce}
      \sup_{t\in(0,T]}|t^{-\delta} f(t,y(t))| < + \infty.
    \end{equation}
  
    \item There is a uniform constant $L>0$ such that for any function
    $y_1$, $y_2$ defined on $(0,T]$ (or on any subset thereof), 
    \begin{equation}
    \label{eq:LipschitzSourcePW}
t^{-\delta} \big| f(t,y_1(t))-f(t,y_2(t)) \big|
    \le L \, t^{-\mu_0} \big| y_1(t)-y_2(t)) \big|,
  \end{equation}
  for each $t$ in $(0,T]$ at which $|t^{-\mu_0}
    y_1(t)|\le s$ and $|t^{-\mu_0} y_2(t)|\le s$. 
  \end{enumerate}
  
  Under these conditions, by an elementary fixed point argument we can prove that there is a time $\widehat T\in (0,T]$ and a {\sl unique smooth solution} 
  $u: (0,\widehat T]\to \R^n$ of \Eqref{eq:FuchsianODEgen} which vanishes at $t=0$ at the following rate: 
  \begin{equation}
    \label{eq:decayestSIVP}
    \sup_{t\in(0,\widehat T]}|t^{-\mu_0} u(t)|\le s.
  \end{equation}
The time of existence $\widehat T$ depends on $\delta$, $\mu_0$, $s$, $L$
  and a constant $d>0$ which we can choose to satisfy
    \begin{equation}
      \label{eq:defd}
      \sup_{t\in(0,T]}|t^{-\delta} f(t,0)|\le d
    \end{equation}
as a consequence of \Eqref{eq:BoundSouce}. Observe that given this, \Eqsref{eq:LipschitzSourcePW} and \eqref{eq:defd} allow us to write \Eqref{eq:BoundSouce} in the more explicit form
\begin{equation}
  \label{eq:BoundSouceref}
\sup_{t\in (0,T]}|t^{-\delta} f(t,y(t))|\le d+L s
\end{equation}
for any smooth function $y:(0,T]\to \R^n$ with
$\sup_{t\in(0,T]}|t^{-\mu_0} y(t)|\le s$.
 
    Given the existence of a solution $u$ with the property \Eqref{eq:decayestSIVP}, we can then integrate \Eqref{eq:FuchsianODEgen} to show the stronger decay rate
  \begin{equation}
    \label{eq:decayestSIVPImpr}
    \sup_{t\in(0,\widehat T]}|t^{-\delta} u(t)|\le C \, (d+Ls),
  \end{equation}
where the constant $C$ depends on $\delta$ and $A$, only, provided $\mu_0$ is larger than the largest real part of all eigenvalues of
    $A$.

\paragraph{Remarks}

The above existence result is standard for ODEs but, interestingly for the purpose the present paper, can also be regarded as a special case of the PDEs result established in \cite{beyer2010b,ames2013a} (with the specified decay statement \eqref{eq:decayestSIVPImpr}). This motivates us to investigate the ODEs problem first before extending our conclusions to the corresponding PDEs problem in the second part of this paper.

Let us continue with a few remarks about this theorem.  
As an illustration, consider the example $f=0$ and $n=1$. The general solution of
\Eqref{eq:FuchsianODEgen} is then $u(t)=u_* t^A$ for an arbitrary $u_*\in \R$. It is clear that there are therefore infinitely many
solutions $u(t)$ with the property $\sup_{t\in(0,T]}|t^{-A} u(t)| < + \infty$. 
This shows that it is crucial that $\mu_0 I -A$ be \emph{strictly positive}
in conditions (i) and (ii) above, for otherwise uniqueness would be lost. 
In this example, the solution $u$ which is uniquely
determined by the condition
$\sup_{t\in(0,\widehat T]}|t^{-\mu_0} u(t)| < + \infty$ with $\mu_0>A$ is
clearly $u\equiv 0$. 
This demonstrates that, given an arbitrary ODE of the form \Eqref{eq:FuchsianODEgen}, the smaller we are allowed to choose $\mu_0$ in the above considerations, the larger the class of functions is among which the solution $u$ with the property \eqref{eq:decayestSIVP} is unique. We emphasize that this does not rule out the existence of other solutions with the property $\sup_{t\in(0,\widehat T]}|t^{-\widetilde\mu_0} u(t)|\le \infty$ for any $\widetilde\mu_0<\mu_0$.


\subsection{Properties of solutions}
\label{sec:properties}

\paragraph{Exponents and transformations}

Let us further discuss the structural conditions for $\mu_0$ and $A$
in the theorem above 
using the example $A=\begin{pmatrix}0 & 1\\0 & 0
\end{pmatrix}
$. It is easy to see that $\mu_0 I-A$ is positive definite if and only if $\mu_0>1/2$. On the other hand all eigenvalues of $A$ are $0$, and therefore $\mu_0$ is larger than the largest real part of all eigenvalues, if $\mu_0>0$. This demonstrates that there is, in general, a non-optimality discrepancy between our two conditions for $\mu_0$ and $A$ (in Section \ref{thm:SIVP}), and this discrepancy disappears only if $A$ is a symmetric matrix. Note in passing that it is possible to refine this theorem  when $A$ has a block diagonal structure, since one can then associate  different constants $\delta$ and $\mu_0$ to each block of $A$. (We will not attempt to make use of this observation here.) 

For some of our following arguments we will rely on transforms from a Fuchsian equation \eqref{eq:FuchsianODEgen} into another Fuchsian equation with different parameters, as follows. 
Let us pick $p>0$ and $k\in\R$, and then set
\begin{equation}
  \label{eq:transf01}
  t=\widetilde t^{p},\quad 
  u(t)=\widetilde t^{-k} \widetilde u(\widetilde t).
\end{equation}
If $u$ is a solution of \Eqref{eq:FuchsianODEgen}, then $\widetilde u$ is a solution of
\begin{equation}
  \label{eq:FuchsianODEgentrans}
  \widetilde t\partial_{\widetilde t} \widetilde u(\widetilde t)-\widetilde A \widetilde u(\widetilde t) = \widetilde f(\widetilde t,\widetilde u(\widetilde t)),
\end{equation}
for
\begin{equation}
  \label{eq:FuchsianODEgentrans2}
  \widetilde A=p A+k I,\quad \widetilde f(\widetilde t,\widetilde y)=p \widetilde t^k f(\widetilde t^p,\widetilde t^{-k}\widetilde y).
\end{equation}
In particular, if \Eqref{eq:FuchsianODEgen} satisfies the condition in Section \ref{thm:SIVP}  for some $\delta$ and $\mu_0$, then \Eqsref{eq:FuchsianODEgentrans}--\eqref{eq:FuchsianODEgentrans2} satisfies the conditions in Section \ref{thm:SIVP}  for 
\begin{equation}
  \label{eq:FuchsianODEgentrans3}
  \widetilde\delta=p\delta+k,\quad \widetilde\mu_0=p\mu_0+k.
\end{equation}

\paragraph{Connection with the \emph{singular initial value problem}}

Let us return to the discussion initiated in \Sectionref{sec:intro}.  Instead of applying the technique of Section \ref{thm:SIVP} to a given equation \Eqref{eq:FuchsianODEgen} directly (which may not be possible if our assumptions are not satisfied), we can often specify an (in principle) free function $u_*(t)$, called the \keyword{leading-order term} and then apply the technique described in Section \ref{thm:SIVP}  to the new Fuchsian ODE \begin{equation}
  \label{eq:FuchsianODEgenrem}
  t\partial_t w(t)-Aw(t) = g(t,w(t)):=f(t,u_0(t)+w(t))-t\partial_t
  u_0(t)+A u_0(t),
\end{equation}
satisfied by the \keyword{remainder} $w=u-u_0$. Interestingly, the additional terms
in $g$ may even compensate potentially ``bad''
terms in the original source term $f$ and therefore allow the setup in Section \ref{thm:SIVP} to apply. In any case, if the conditions in Section \ref{thm:SIVP} hold to this last equation,
\Eqref{eq:FuchsianODEgenrem} has a solution $w$ uniquely determined by the
condition $\sup_{t\in(0,\widehat T]}|t^{-\mu_0} w(t)| < + \infty$. If the
exponents $\mu_0$ and $\delta$ can be picked up to be sufficiently large, the remainder $w$ can be
interpreted as a \emph{higher order} contribution to the solution $u$ with
\keyword{leading-order term} $u_0$ at $t=0$.  

Clearly, the solution
$w(t)$ of \Eqref{eq:FuchsianODEgenrem}
depends in general on the choice of $u_0$. One can therefore
interpret the freedom of specifying $u_0$  as the freedom to specify 
\emph{asymptotic data} for the
solution $u$ of \Eqref{eq:FuchsianODEgen}. In order to distinguish this from the \emph{Cauchy problem} (or regular \keyword{ initial value problem}), where the freedom is to
choose the value of the unknown $u$ at some \emph{regularity time} $t_*>0$, one
refers to this as the \keyword{singular initial value problem}. The singular initial value problem differs significantly
from the Cauchy problem. At least in the case $n=1$ this is not directly obvious since it seems that we can always get rid of the main singular term in the Fuchsian equation by a transformation of the form \Eqref{eq:transf01}. In fact, if $\delta$ is an arbitrary constant with
$\delta>A$ in the case $n=1$, we pick 
\begin{equation}
  \label{eq:RegCauchyProblTrafo}
  p=\frac{1}{\delta-A},\quad k=-\frac{A}{\delta-A},
\end{equation}
then 
\begin{equation}
  \label{eq:RIVP1N}
  \partial_{{\widetilde t}}\widetilde u({\widetilde t})
  =F({\widetilde t},{\widetilde u}({\widetilde t}))
\end{equation}
where
\begin{equation}
  \label{eq:RIVP2N}
  F({\widetilde t},\widetilde y)=\widetilde t^{-1}\widetilde f({\widetilde t},\widetilde y)
  =\frac 1{\delta-A} {{\widetilde t}}^{-\delta/(\delta-A)} 
  f({{\widetilde t}}^{1/(\delta-A)}, {{\widetilde t}}^{A/(\delta-A)} \widetilde y).
\end{equation}

At a first glance, it appears that finding a solution of
\Eqref{eq:FuchsianODEgen} with the property
$\sup_{t\in(0,T]}|t^{-\mu_0} u(t)| < + \infty$ for some
$\mu_0\in (A,\delta)$ as Section \ref{thm:SIVP}  might be
equivalent to solving the \emph{regular} Cauchy problem of
\Eqsref{eq:RIVP1N} and \eqref{eq:RIVP2N} with initial data
\begin{equation}
  \label{eq:regulardata1N}
  {\widetilde u}(0)=0.
\end{equation}
This however is in general false. The conditions for well-posedness of
the Cauchy problem (which can be found in most text books about
ordinary differential equations) of \Eqsref{eq:RIVP1N},
\eqref{eq:RIVP2N} and \eqref{eq:regulardata1N} translate into the
following two conditions for the original function $f$ (restricting to $n=1$ as mentioned
above): There is a constant $\delta>A$ such that:
\begin{enumerate}[label=(\Roman{*}),leftmargin=*]
\item  For some $\tildeT,\widetilde s>0$ the following function
  \begin{equation}
    \label{eq:regularcond1}
    (t, y)\mapsto {t}^{-\delta} f(t, {t}^{A}y)
    \text{ 
  can be extended continuously to the domain 
  $\widetilde D=[0,\widetilde  T]\times[-\widetilde s,\widetilde s]$.} 
  \end{equation}

\item There is a constant $L>0$ such that 
  \begin{equation}
    \label{eq:regularcond2}
    |{t}^{-\delta} (f(t, {t}^{A} y_1)-f(t, {t}^{A}y_2))|
    \le L| y_1- y_2| \text{   for all
  $(t, y_1), (t, y_2)\in \widetilde D$.}
  \end{equation}

\end{enumerate}
Even though these two conditions look similar to those in Section \ref{thm:SIVP}, it is clear that the setup therein is more
general. For example, the case $A=0$ and $f(t,y)=y^2$ is covered by Section \ref{thm:SIVP}, but the function
${t}^{-\delta} f(t, y)={t}^{-\delta} y^2$ fails
condition (I) for every $\delta>A=0$. The singular initial
value problem of \Eqref{eq:FuchsianODEgen} in general can 
not be transformed to a (regular) Cauchy problem for
\Eqsref{eq:RIVP1N}, \eqref{eq:RIVP2N} and \eqref{eq:regulardata1N} via
the transformation \Eqref{eq:transf01} with \Eqref{eq:RegCauchyProblTrafo}. 
This does
 not rule out the possibility that with a suitable (for
instance nonlinear) transformation of some particular Fuchsian
equation  the singular initial value problem can be turned into
a regular Cauchy problem; an example is given
in \Sectionref{sec:ODEtestproblem}.


\subsection{Approximation of Fuchsian equations}

Our main objective is to \emph{accurately calculate} solutions of the singular initial value problem, possibly for partial differential equations but we continue by restricting attention to ordinary differential equations.   
We claim that the solutions to the singular problem can be
\emph{approximated} by solutions to regular Cauchy problems with vanishing data imposed at arbitrary $t=t_*>0$. The statement below is the ODE version of the PDE theory established in \cite{beyer2010b, ames2013a}. 
 
 Let us introduce the following notation.  Given an arbitrary $t_*\in (0,T]$, we define a time $T_{t_*}\in (0,T]$ and a function $v_{t_*}: (0, T_{t_*})\to \R^n$ as follows. Consider the unique solution to the regular Cauchy problem of \Eqref{eq:FuchsianODEgen} with vanishing initial data at $t=t_*$. Then, $T_{t_*}$ is determined by the requirement that $[t_*,T_{t_*})$ is the maximal future existence interval in $[t_*,T]$ of this solution. 
 We then set $v_{t_*}(t)=0$ for all $t\in (0, t_*]$ while $v_{t_*}(t)$ is defined to coincide with the solution of the Cauchy problem for $t\in [t_*,T_{t_*})$.
 
\begin{theorem}
  \label{thm:approx}
Consider the singular initial value problem for some $A$ and $f$, $T$, $s$, $\delta$, $\mu_0$ and $L$ satisfying the condition in Section \ref{thm:SIVP}.   Then the following properties hold. 

\begin{enumerate}[label=(\roman{*}),leftmargin=*]
\item There exists $\tildeT\in (0,T]$ such that, for 
  every $t_*\in (0, \tildeT]$, we have $\tildeT\le T_{t_*}$ and
  \begin{equation}
    \label{eq:CPsolest}
    \sup_{t\in (0,\tildeT]} |t^{-\mu_0} v_{t_*}(t)|\le s.
  \end{equation}
  This time $\tildeT$ depends on $\delta$, $\mu_0$, $s$, $L$ and $d$. 
 If $\mu_0$ is larger than the largest real part of all eigenvalues of
    $A$, then
    \begin{equation}
    \label{eq:CPsolest2}
    \sup_{t\in(0,\tildeT]}|t^{-\delta} v_{t_*} (t)|\le \widetilde C (d+L s),
  \end{equation}
  where $\widetilde C>0$ only depends on $\delta$ and $A$.
\item There is a constant $C>0$, such that, for every $t_*\in (0, \tildeT]$ and for
  any $\lambda$ larger than the largest real part of all eigenvalues
  of $A$, we have
  \begin{equation}
    \label{eq:approximationest}    
    \sup_{t\in (0,\tildeT]}|t^{-\lambda} (u(t)-v_{t_*}(t))|\le C
    t_*^{\delta-\lambda}.
  \end{equation}  
Here $u: (0,T]\to \R^n$ is the solution of
  \Eqref{eq:FuchsianODEgen} asserted in Section \ref{thm:SIVP} (without
  loss of generality we assume that $T=\widehat T$).  The constant $C$ only depends on $\delta$, $\mu_0$, $T$, $s$, $L$ and $d$.
\end{enumerate}
\end{theorem}

The item~(ii) above states that the solution $u$ of the singular initial value problem can be  approximated by solutions $v_{t_*}$ of the regular Cauchy problem with vanishing data at any arbitrary \emph{regularity
  time} $t_*>0$. The closer $t_*$ is to zero, the better the
approximation, as stated in \Eqref{eq:approximationest} which provides us with an estimate for the continuum approximation error introduced in \Sectionref{sec:intro}. 
Part~(i) states that these approximate solutions \emph{exist on a common time interval} $(0,\tildeT]$ irrespective of the initial time $t_*$. This guarantees that the approximations are valid uniformly in time.


\section{Weighted error estimates for Fuchsian equations} 
\label{sec:errorestimates}

\subsection{Aim for the rest of this paper} 

We propose to numerically approximate the solutions $u=u(t)$ to the \emph{singular} initial value problem by 
a sequence of solutions $v=v_{t_*}(t)$ to the
\emph{regular} Cauchy problem (with vanishing data at $t=t_*$) where the sequence $t_* \to 0$ as suggested by \Theoremref{thm:approx}. In what follows
we use the same terminology of the two main errors of interest, that is,  the
\emph{numerical approximation error} and the \emph{continuum
  approximation error}. Both errors combined are 
referred to as the \emph{total approximation error}.

After we have introduced a family of one-step methods in \Sectionref{sec:onestepmethods} below, we are going to select a
particular scheme and derive quantitative estimates for the associated
numerical approximation error. For simplicity, we focus on the class of singular Fuchsian
equations \Eqref{eq:FuchsianODEgen}, and we recall that \Theoremref{thm:approx} provides an estimate for the continuum approximation error. Standard textbook estimates do not apply here since the solutions of interest, in general, are not smooth up to the singularity
$t=0$. After deriving quantitative bounds for the numerical error, we will then study how this error can be balanced
with the continuum approximation error given by \Eqref{eq:approximationest}. Later in \Sectionref{sec:Numtestproblem} we will validate our theoretical estimates numerically on a test
problem and finally we will discuss our main application
in \Sectionref{sec:mainnumapplications}.


\subsection{A one-step method}
\label{sec:onestepmethods}

In what follows, some constants $T>t_*>0$ and a positive
integer $N$ are fixed. Given an arbitrary sequence of positive reals
$(h_i)_{i=0,\ldots,N-1}$ we set
\begin{equation}
  \label{eq:timegrid}
  t_{i+1}=t_i +h_i, \qquad i=0,\ldots,N-1,\quad t_0=t_*,
\end{equation}
chosen so that $t_N=T$.  It is convenient to write this as
\begin{equation}
  \label{eq:timegridalpha}
  t_{i+1}=t_i(1+\alpha_i),\quad \alpha_i:=h_i/t_i, \qquad i=0,\ldots,N-1. 
\end{equation}
A general one-step method (cf for instance \cite{gautschi2012}) to
numerically approximate the forward Cauchy problem of an ODE with vanishing Cauchy 
data at the initial time $t=t_*$ can be expressed with a smooth
$n$-vector valued function $\Phi(t,y,\alpha)$ (depending on the
ODE and scheme under consideration),  as follows
\begin{equation}
  \label{eq:singlestepscheme}
  y_{i+1}=y_i+ \Phi(t_i, y_i; \alpha_i),\quad y_0=0,\qquad i=0,\ldots, N-1.
\end{equation}
The numerical solution is thus given by a sequence of $n$-dimensional vectors $(y_i)_{i=0,\ldots,N}$.  Since $\Phi$ is a smooth
$n$-vector-valued function, we can always find a smooth
$n\times n$-matrix-valued map $K(t, y, \widetilde y;\alpha)$ satisfying
\begin{equation}
  \label{eq:DefK}
  \Phi(t,y;\alpha)-\Phi(t,\widetilde y; \alpha)=K(t,y,\widetilde y; \alpha) \, (y-\widetilde y). 
\end{equation}
This map will be useful for our analysis.
On the other hand, the \keyword{truncation
  error}, by definition, is the function
\begin{equation}
  \label{eq:truncationerror}
  \Theta(t,y;\alpha)=\Phi(t,y;\alpha)+y -\widetilde v(t(1+\alpha)),
\end{equation}
where $\widetilde v(t(1+\alpha))$ is the value of
the solution $\widetilde v$ of the given ODE at time $t(1+\alpha)$ determined by the Cauchy data $\widetilde v(t)=y$.

Now let\footnote{This function was denoted by $v_{t_*}$ in \Theoremref{thm:approx}.}
$v$ be the (exact) solution of the Cauchy problem of the given
differential equation with vanishing Cauchy  data at $t=t_*$.  We
write $v_i=v(t_i)$ using \Eqref{eq:timegrid}. For the sequence $(y_i)_{i=0,\ldots,N}$ given by \Eqref{eq:singlestepscheme} we
find (for $i=0,\ldots,N-1$) 
  \begin{align*}
    v_{i+1}-y_{i+1}
    &=(v_{i}-y_{i})-\Phi(t_i, y_i; \alpha_i)+(v_{i+1}-v_i)\\
    &=v_{i}-y_{i}+\left(\Phi(t_i, v_i; \alpha_i)-\Phi(t_i, y_i; \alpha_i)\right)
      +\left(v_{i+1}-v_i-\Phi(t_i, v_i; \alpha_i)\right)\\
    &=v_{i}-y_{i}+K(t_i,v_i,y_i;\alpha_i)(v_i-y_i)-\Theta(t_i,v_i;\alpha_i),
  \end{align*}
which yields   
\begin{equation}
  \label{eq:absnumerror}
  v_{i+1}-y_{i+1}=\left(I+K_i\right) (v_i-y_i) -\Theta_i,
\end{equation}
where $I$ is the $n\times n$-identity matrix, and where  the short-hand notation
\begin{equation}
  \label{eq:notation}
  K_i=K(t_i,v_i,y_i;\alpha_i),\qquad \Theta_i = \Theta(t_i,v_i;\alpha_i)
\end{equation}
has been introduced.

Clearly, $|v_{i+1}-y_{i+1}|$ is the absolute {numerical
  approximation error} at time $t_{i+1}$. In the following we wish to estimate this error
relative to some \emph{weights}, i.e., some sequence
$(w_i)_{i=0,\ldots,N}$ of positive reals. Eventually we will
restrict to weights of the form $w_i=t_i^{-\lambda}$ for exponents
$\lambda$ in analogy to the weight $t^{-\lambda}$ in
\eqref{eq:approximationest}, but for now we allow $w_i$ to be arbitrary
positive reals. Given such weights $(w_i)_{i=0,\ldots,N}$ we define
\begin{equation}
  \label{eq:defweight}
  \omega_i=\frac{w_{i+1}}{w_i},\quad i=0,\ldots,N-1.
\end{equation}
This allows us to write \Eqref{eq:absnumerror} as
\begin{equation} 
  w_{i+1}(v_{i+1}-y_{i+1})
  =\left(1+K_i\right) \omega_i w_{i} (v_i-y_i) -\omega_i w_{i} \Theta_i,
\end{equation}
and therefore
\begin{equation}
  \label{eq:relnumerror}
  |w_{i+1}(v_{i+1}-y_{i+1})|
  \le\left|(1+K_i) \omega_i\right| |w_{i} (v_i-y_i)| +|\omega_i w_{i} \Theta_i|.
\end{equation}
Recall that we denote by $|\cdot|$ the Euclidean norm for
$n$-dimensional vectors or the standard matrix norm.

From the initial condition $v_{0}=y_{0}=0$, we get inductively (using
the convention $\prod_{j=i}^{i-1} ( \ldots ) =1$)
\begin{equation}
  \label{eq:errorestimate5555}
  |w_{0}(v_{0}-y_{0})|=0,\quad
  |w_{i}(v_{i}-y_{i})|\le \sum_{l=0}^{i-1} |\omega_l w_{l} \Theta_l|
  \prod_{j=l+1}^{i-1} \left|(1+K_j) \omega_j\right|,\quad i=1,\ldots,N.
\end{equation}
This leads us to the following \emph{fundamental numerical error estimate} 
\begin{equation}
  \label{eq:errorestimate39842}
  \sup_{i=0,\ldots,N}|w_{i}(v_{i}-y_{i})|\le \sum_{l=0}^{N-1} |\omega_l w_{l} \Theta_l|
  \prod_{j=l+1}^{N-1} \left|(1+K_j) \omega_j\right|.
\end{equation}

Before we proceed further, we raise here two additional observations.  First of all, it will be crucial
to obtain estimates which are {\sl independent of $N$}. For this reason, we wish to estimate sums over discrete time steps by
integrals over a continuous time variable.  The following
relationship between sums and integrals turns out to be very useful: given $\varepsilon\not= 0$ and some integers $i=0,\ldots,N$, we have
  \begin{equation}
    \label{eq:integralestimate}
    {t_i^\varepsilon-t_*^\varepsilon}=\varepsilon\int_{t_*}^{t_i}  t^{\varepsilon-1} dt
    =\varepsilon\sum_{l=0}^{i-1} 
    \int_{t_l}^{t_{l+1}} t^{\varepsilon-1} dt 
      =\sum_{l=0}^{i-1}
      t_l^{\varepsilon}\left((1+\alpha_l)^\varepsilon-1\right),
    \end{equation}
where we used the definition of $\alpha_l$ in \Eqref{eq:timegridalpha}.

Second, we are interested in practically useful restrictions on the
time step sizes. Here we demand that there is a constant $\eta\in [0,1)$ and a constant $H_0>0$ such that
(for $i=0,\ldots,N-1$) 
\begin{equation}
  \label{eq:alpharestr}
  0<\alpha_i\le H_0 t_i^{\eta-1}. 
\end{equation}
Using 
\Eqref{eq:integralestimate} we deduce that 
\[t_i^{1-\eta}-t_*^{1-\eta}=
\sum_{l=0}^{i-1}
      t_l^{1-\eta}\left((1+\alpha_l)^{1-\eta}-1\right).\]
Since $0<1-\eta\le 1$, it is elementary to check that 
$(1+\alpha_l)^{1-\eta}-1\le
(1-\eta)\alpha_l$, and therefore
\[t_i\le \left(t_*^{1-\eta}+(1-\eta) H_0 i\right)^{1/(1-\eta)},
\qquad i=0,\ldots,N. 
\] 
 We
conclude that our sequence of times arising in \Eqref{eq:timegrid} grows at
most like $i^{1/(1-\eta)}$.

Consequently, \Eqref{eq:alpharestr} with $\eta\in [0,1)$
and $H_0>0$ is therefore a relevant restriction on the growth of the time lengths. 
In contrast, observe here that the limit $\eta \to 1$ corresponds to \emph{exponentially
growing} time lengths, which was used by the authors in \cite{beyer2010b}. However, we will show below that such an exponential time stepping is not beneficial for numerical accuracy and may even be prohibitive in practice.


\subsection{Second-order Runge-Kutta method}
\label{sec:mainanalyticalresult}

We now derive uniform quantitative estimates for the numerical error in the case of 
\keyword{second-order Runge-Kutta scheme} (RK2). This scheme is well-known to enjoy numerical stability and provide a good efficiency in practice. On the other hand, our experience with Fuchsian equations suggests that the RK2 scheme is indeed
the best compromise between accuracy and efficiency. Higher order schemes (such as the $4$th order Runge-Kutta
scheme) would only increase the numerical work without always reducing the
total approximation error in the regime that the error is saturated by the continuum approximation error.  
Moreover, the following
analysis of the RK2 scheme is analytically straightforward and 
 serves as a blueprint for more complex schemes.

RK2 is a single step scheme of the form \Eqref{eq:singlestepscheme} with
\begin{equation}
  \label{eq:RK2}
  \begin{split}
    \Phi(t,y; \alpha)= \frac {2\alpha}{2+\alpha} \Biggl(
    &
    A\left(1+\frac{A\alpha}2\right)y+\frac {A\alpha} 2  f(t,y)
    +
    f\left(\left(1+\frac \alpha 2\right)t,
      \left(1+\frac{A\alpha}2\right)y+\frac \alpha 2  f(t,y)\right)
    \Biggr), 
  \end{split}
\end{equation}  
which we now apply to
\Eqref{eq:FuchsianODEgen}.
A first observation is that the RK2 scheme is \emph{not invariant under
transformations} of the form \Eqref{eq:transf01}.
It is therefore an interesting question whether every equation of the form \Eqref{eq:FuchsianODEgen} allows for a transformation
which somehow minimizes the numerical approximation error; we explore this question (among other things) in \Sectionref{sec:discparam} below.

In order to be able to perform our analysis of the RK2 scheme, we assume now that 
the constant $H_0$ in \Eqref{eq:alpharestr} can be written as
\begin{equation}
  \label{eq:boundednessalpha}
  H_0=H_1 t_*^{1-\eta+\beta},
\end{equation}
for some other constants $H_1>0$ and $\beta\ge 0$.
Most significantly this implies that the sequence $(\alpha_i)_{i=0,\ldots,N-1}$ is
\emph{uniformly bounded} since
\begin{equation}
  \label{eq:alpharestrbounded} 
  0<\alpha_i\le H_1 t_*^{1-\eta+\beta} t_i^{\eta-1}\le H_1 t_*^\beta,
\end{equation}
for each $i=0,\ldots,N-1$. Allowing $H_1$ or $t_*$ to be small, without loss of generality  we can  assume 
that all sequence elements $\alpha_i$ are arbitrarily small uniformly over the
whole time interval of interest. 
This assumption is crucial for the proof of our main theorem below. In any case, it is clear that \Eqref{eq:boundednessalpha} with $\beta\ge 0$ is a genuine restriction, and it is at this stage not clear how this affects practical
computations; see \Sectionref{sec:numnegbeta}.

\begin{theorem}[Global numerical approximation error for Fuchsian equations: the RK2 scheme]
  \label{theorem:RK2}
  Pick $\beta\ge 0$, $\eta\in [0,1)$, $T\in (0,1]$, $s>0$, and $A$ and
  $f$ as in Section 2. Suppose there are constants $\delta$, $\mu_0$ and $L$
  such that the conditions in Section \ref{thm:SIVP} are satisfied, and,
  in addition that $\mu_0$ is larger than the largest real part of all
  eigenvalues of $A$ and that $\delta>|A|$. Moreover assume that 
  there is a constant $d>0$ such that
  \bse
  \begin{equation} 
    \sup_{t\in (0,T]}\left|t^{-(\delta-l_1-|l_2|\mu_0)}\partial_t^{l_1}\partial_u^{l_2}f(t,0)\right|
    \le d
  \end{equation}
  for every 
  non-negative integer $l_1$ and $n$-dimensional multiindex $l_2$
  with $l_1+|l_2|\le 3$. Finally suppose that for
  any functions $z_1$ and $z_2$ defined on a subset of $(0,T]$ and every
  $l_1$ and $l_2$ as above,  
  \begin{equation}
    \label{eq:LipschitzSourcePWDer}
    \left|t^{-(\delta-l_1-|l_2|\mu_0)}\left((\partial_t^{l_1}\partial_u^{l_2}f)(t,z_1(t))-(\partial_t^{l_1}\partial_u^{l_2}f)(t,z_2(t))\right)\right|
    \le L |t^{-\mu_0}(z_1(t)-z_2(t))|,
  \end{equation}
  for all values of $t$ in $(0,T]$ for which
  $|t^{-\mu_0} z_1(t)|\le s$ and $|t^{-\mu_0} z_2(t)|\le s$. Then the following properties hold. 
  \ese
  
  \begin{itemize}
  
  \item {\it Global existence and boundedness.}   
  Provided $T$ is sufficiently small, one can pick an arbitrary integer $N>0$, an arbitrary $t_*\in (0,T)$ and an arbitrary sequence $(\alpha_i)_{i=0,\ldots,N-1}$, which satisfies
  \Eqsref{eq:alpharestr} and
  \eqref{eq:boundednessalpha} for a sufficiently
  small constant $H_1>0$, so that the sequence
  $(y_i)_{i=0,\ldots,N}$ defined by \Eqsref{eq:singlestepscheme}, \eqref{eq:RK2} and
  \eqref{eq:timegridalpha} is well-defined and 
$\sup_{i=0,\ldots, N}|t_{i}^{-\mu_0} y_{i}|\le s/2.$
   
   \item {\it Error estimate.} For any $\lambda\in (|A|,\delta)$, the numerical approximation error is bounded as follows
   \bse
     \label{eq:mainerrorestimate00}
   \begin{equation}
  \label{eq:mainerrorestimate}
  \sup_{i=0,\ldots,N}|t_{i}^{-\lambda}(v_{t_*} (t_{i})-y_{i})|\le \frac
  C{|\delta-\lambda+2(\eta-1)|} 
  H_1^2 t_*^{\sigma_{num}},
\end{equation}
with
\begin{equation}
  \label{eq:mainerrorestimateexp}
  \sigma_{num}=2\beta+\min\{2(1-\eta), \delta-\lambda\}, 
\end{equation}
for some constant $C>0$, which only depends on $d$, $L$, $s$,
$\delta$, $A$, $\mu_0$ and $H_1$. 
Here,
$v_{t_*}$ is the function introduced in
\Theoremref{thm:approx} (supposing without loss of generality that $\tildeT=T$).
\ese

\end{itemize}
\end{theorem}

Before we give the proof of \Theoremref{theorem:RK2} in \Sectionref{sec:proof}, let us discuss here some consequences. First of all, the restriction $\delta>|A|$ is probably not  optimal. In general  the condition $\delta>\mu_0$ and that $\mu_0$ is larger than the largest real part of all eigenvalues of $A$ should be sufficient. We expect the condition $\delta>|A|$ to be especially restrictive when $A$ has eigenvalues with very large negative real parts. In this case, we can however apply the transformation \Eqref{eq:transf01} with some sufficiently large positive $k$ and work with the transformed system whose matrix $\widetilde A$ in \Eqref{eq:FuchsianODEgentrans2} does not have such eigenvalues.
First of all,  
the technical restriction $\delta>|A|$ need not be optimal when $A$ admits negative eigenvalues. 
Hence, we should always try to apply a transformation like \Eqref{eq:transf01} in order to make $A$ positive definite as a consequence of \Eqref{eq:FuchsianODEgentrans2}, and only then apply the above theorem.

Our theorem provides some essential information on the numerical evolution of solutions to Fuchsian equations. 
It implies that, as long as we choose $T$ and $H_1$ sufficiently small (and the other more technical conditions are met), the numerical evolution
extend (and enjoys uniform bounds) to the common time $T$. This is irrespectively of the choice of the initial time $t_*$
and the number of time steps $N$. This notion of \emph{global
  existence for the numerical scheme} is the core of our strategy for accurately approximating (in the limit $t_* \to  0$) 
  the singular Cauchy problem by  numerical solutions to the regular Cauchy problem. Given this property, our main result is the estimate \Eqsref{eq:mainerrorestimate00} for the numerical approximation error.


\subsection{Our strategy concerning the discretization parameters}
\label{sec:discparam}

\paragraph{Asymptotically balanced discretizations}

\Eqsref{eq:mainerrorestimate00} is our fundamental estimate for the numerical approximation error which depends upon the main parameters $H_1$, $t_*$ and $\lambda$. Of special importance is the dependence on $t_*$, since we shall mostly consider the limit $t_*\to  0$ for fixed $H_1$ and $\lambda$. The bigger the exponent $\sigma_{num}$ is, the faster the numerical approximation error approaches zero in this limit. Observe that the closer $\eta$ is to $1$ 
(the limit case being the exponential time stepping; see above),
the smaller this exponent is. In general, exponential time stepping does therefore not lead to a good numerical strategy.

The total
approximation error is a combination of the numerical
approximation error estimated by \Eqsref{eq:mainerrorestimate00} and the continuum approximation error
estimated by \Eqref{eq:approximationest}. Considering $H_1$ and $\lambda$ as fixed, both estimates bound the respective errors by a power of $t_*$, in the first case by $t_*^{\sigma_{num}}$, and  in the second case by $t_*^{\sigma_{cont}}$. Here, we have 
\begin{equation}
  \label{eq:mainerrorestimateexp2}
  \sigma_{cont}=\delta-\lambda, 
\end{equation}
and we immediately conclude that 
\begin{equation}
\label{eq:sigmaest}
\sigma=\min\{\sigma_{num},\sigma_{cont}\}
=
\begin{cases}
  \sigma_{cont}=\delta-\lambda & \text{ if }
  \beta\ge \max\{(\delta-\lambda)/2-(1-\eta),0\},\\
  \sigma_{num}=2\beta+2(1-\eta) & \text{ if }
  \beta\le \max\{(\delta-\lambda)/2-(1-\eta),0\},
\end{cases}
\end{equation}
where the last case is possible (under the restriction $\beta\ge 0$) only if $\delta-\lambda\ge 2(1-\eta)$.
We say that \textbf{the continuum and the numerical approximation errors are asymptotically balanced} if $\sigma_{num}=\sigma_{cont}=\sigma$. This
is therefore the case if and only if 
\begin{equation}
  \label{eq:balancedbeta}
  \beta=\max\big\{ (\delta-\lambda)/2-(1-\eta),0\big\}.
\end{equation}
In the applications, one should, in principle, strive for asymptotic balance.

Choosing $\beta$  larger than that in \Eqref{eq:balancedbeta} would mean that we are doing too much numerical work (in particular, the number of time steps given by \Eqref{eq:alpharestr} with \Eqref{eq:boundednessalpha}) is unnecessarily high) without improving the accuracy of the approximation since the total approximation error is asymptotically saturated  by the continuum approximation error. Choosing $\beta$ smaller than that in \Eqref{eq:balancedbeta} would mean that the approximation is ``not as good as it could be'' since the numerical resolution is too small.

  \paragraph{Cost of computation}
  
The main benefit of choosing $\beta$ smaller than that in \Eqref{eq:balancedbeta} however is that we obtain an approximation with a relatively small number of time steps. Let us assume here that in addition to the upper bound \Eqref{eq:alpharestr} with \Eqref{eq:boundednessalpha}) for the time steps, there is also a lower bound
of the form 
\[\alpha_i\ge H_2 t_*^{1-\eta+\beta} t_i^{\eta-1},\]
for some $0<H_2\le H_1$. A very rough estimate for the number $N$ of time steps, which take us from the initial time $t_*$ to the end time $T$ is therefore
\begin{equation}
  \label{eq:estimateruntime}
N\le \frac{T-t_*}{H_2 t_*^{1+\beta}}=C t_*^{-(1+\beta)}.
\end{equation}
Observe that this estimate is true for all $\eta\in [0,1)$,
but only  optimal in the case $\eta=0$, i.e., the case where the time step sizes are bounded by a constant (in this paper we mostly  focus on constant time step sizes for simplicity and therefore mostly choose $\eta=0$). To get a better estimate for $N$ when $\eta\not=0$, we rewrite the inequality
\[t_{i+1}-t_i=h_i\ge H_2 t_*^{1-\eta+\beta} t_i^{\eta},\]
as
\[\frac{t_{i+1}-t_i}{\zeta_{i+1}-\zeta_i}\ge  t_i^{\eta},\]
where $\zeta_i=H_2 t_*^{1-\eta+\beta} i$. If $H_2 t_*^{1-\eta+\beta}\ll 1$, we can interpret $\zeta$ as a continuous variable and the quantity $t$ as a function of $\zeta$ so that the inequality becomes
\[t'(\zeta)\ge  t^{\eta}(\zeta).\]
From this we conclude the improved estimate that
\begin{equation}
  \label{eq:estimateruntimeimpr} 
  N\le C t_*^{-(1+\beta-\eta)}.
\end{equation}
For simplicity we shall now however only work with the estimate \Eqref{eq:estimateruntime}.
Given that estimate and the estimate $E\le C t_*^\sigma$ for the total approximation error with $\sigma$ given by \Eqref{eq:sigmaest} we conclude that
\[E N^{\sigma/(1+\beta)}\le C.\] 
This relationship between the total approximation error $E$ and the numerical work $N$ can be interpreted as a statement about the \emph{asymptotic efficiency}.
Our
approximation of solutions of the singular initial value problem is thus more \emph{asymptotically efficient}, the larger
the following \textbf{asymptotic efficiency exponent} is: 
\begin{equation*} 
\frac{\sigma}{1+\beta}
=
\begin{cases}
  \frac{\delta-\lambda}{1+\beta} & \text{ if }
  \beta\ge \max\{(\delta-\lambda)/2-(1-\eta),0\},\\
  2-2 \frac{\eta}{1+\beta}  & \text{ if }
  \beta\le \max\{(\delta-\lambda)/2-(1-\eta),0\}. 
\end{cases}
\end{equation*} 

\paragraph{The relevant regime}

Let us restrict now to the case $\eta=0$. 
The optimal asymptotic efficiency \emph{exponent $2$ is achieved by setting} $\beta=0$ when
$\delta-\lambda<2$, or, by choosing an
\emph{arbitrary} value $\beta$ in the interval
$[0, (\delta-\lambda)/2-1]$ if $\delta-\lambda\ge 2$. Recall that in the first case, the approximation is also asymptotically balanced, while in the second case, this is only true if $\beta$ has the maximal value $(\delta-\lambda)/2-1$ in that interval. Even though any other value for $\beta$ in that interval does not lead to asymptotic balance, the cost in loss of accuracy associated with this is cancelled by the benefit of decreased numerical work.
We have therefore found the following interesting result (in the case $\eta=0$). 
\textbf{We
  can always choose $\beta=0$ in order to achieve optimal asymptotic efficiency}. 
  From the practical point of view this is good news since the choice $\beta=\eta=0$ is also the easiest one to implement as we discuss below. 

\paragraph{Applying a transformation of the Fuchsian equation}

Now we address the question how these results are affected by transformations of the form \Eqref{eq:transf01}. In particular can we always choose the parameters $p$ and $k$ there to map our equation into the regime of maximal efficiency? The answer is no. Suppose that everything has been chosen so that the conditions for \Theoremref{theorem:RK2} are satisfied for an arbitrary given Fuchsian equation \eqref{eq:FuchsianODEgen}. Now pick $p$ and $k$ in \Eqref{eq:transf01} and consider the transformed Fuchsian equation \eqref{eq:FuchsianODEgentrans} with \Eqref{eq:FuchsianODEgentrans2}. We have seen that we should pick $\widetilde\delta$ and $\widetilde\mu_0$ according to \Eqref{eq:FuchsianODEgentrans3}. Given any $\lambda$ to measure the error for the approximate solutions of \eqref{eq:FuchsianODEgen}, the definition of the weighted norm together with \Eqref{eq:transf01} implies that we should choose
\[\widetilde\lambda=p\lambda+k\]
to measure the errors for the transformed problem. It therefore follows that
\[\widetilde\delta-\widetilde\lambda=p(\delta-\lambda).\]
The definition of $\alpha$ in \Eqref{eq:timegridalpha} implies (at least in the limit of small time steps when $\alpha=h/t\approx \frac 1t \frac{dt}{d\widetilde t}d\widetilde t=p\widetilde h/\widetilde t=p\widetilde\alpha$, which is the case when $H_1$ or $t_*$ are very small), that
 the above condition 
\[H_2 t_*^{1-\eta+\beta} t_i^{\eta-1}\le \alpha_i\le H_1 t_*^{1-\eta+\beta} t_i^{\eta-1}\]
is satisfied if and only if 
\[\widetilde H_2 \widetilde t_*^{1-\widetilde\eta+\widetilde\beta} t_i^{\widetilde\eta-1}\le \widetilde\alpha_i\le \widetilde H_1 \widetilde t_*^{1-\widetilde\eta+\widetilde\beta} \widetilde t_i^{\widetilde\eta-1}\]
holds
for
\begin{equation}
  \label{eq:transfparameters}
  \widetilde H_1=H_1/p,\quad \widetilde H_2=H_2/p,\quad 1-\widetilde\eta=p(1-\eta),\quad \widetilde\beta=p\beta.
\end{equation}

\paragraph{A practical conclusion}

We conclude from \Eqref{eq:sigmaest} that
\[\widetilde\sigma=p\sigma,\]
and that the approximation errors are therefore asymptotically balanced for the original problem of \Eqref{eq:FuchsianODEgen} if and only if they are asymptotically balanced for the transformed problem according to \Eqref{eq:balancedbeta}. Using the estimate \eqref{eq:estimateruntimeimpr} (which is better than the estimate \eqref{eq:estimateruntime} in the case $\eta>0$; but it only holds in the same limit of very small time steps assumed in our discussion here), we in particular see that the asymptotic efficiency exponent is invariant, i.e.,
\[\frac{\sigma}{1-\eta+\beta}=\frac{\widetilde\sigma}{1-\widetilde\eta+\widetilde\beta}.\]
We have therefore found that \textbf{even though the RK2-scheme is not invariant under transformations of the form \Eqref{eq:transf01}, asymptotic balance and asymptotic efficiency are invariant}. It is also interesting to observe that the choice $\eta=0$, which we are considering in most parts of this paper here, is not really a restriction of generality: Suppose we would work in a situation where $\eta\in (0,1)$. Then, by means of a transformation \Eqref{eq:transf01} with $p=1/(1-\eta)$, we could map the problem to a situation with $\widetilde\eta=0$, see \Eqref{eq:transfparameters}.

\paragraph{Constant factors}

Our last discussion is now about the fact that all error estimates above include undetermined constants $C$. This means that even if we set things up so that the approximation is asymptotically balanced and/or has optimal asymptotic efficiency, the numerical approximation error and the continuum approximation error may differ by many orders of magnitude in size. 
In order to balance not only the asymptotic decay rates of these two errors
(given by the
exponents $\sigma_{num}$ and $\sigma_{cont}$), but also their
absolute sizes, we propose the following practical algorithm. Suppose we want to approximate the solution of the singular initial value problem by calculating a sequence of numerical solution of the forward Cauchy problem determined by a monotonic sequence of initial times $t_*$ for fixed $H_1$ and $\lambda$ as above. Then
\begin{enumerate}
\item Pick the largest value of \(t_*\) of interest.
\item Given this value of $t_*$ (and some fixed value of $\lambda$), choose a decreasing sequence of positive $H_1$-values, calculate the numerical approximations and then estimate each total approximation error (we discuss how to do this below). So long as the total approximation errors are dominated by the numerical approximation errors, \Eqref{eq:mainerrorestimate} guarantees that the sequence of total approximation errors decreases as $H_1$ decreases. 
\item In this way, find the value for $H_1$ where the total approximation error stops decreasing as $H_1$ decreases. This is then the point where the numerical and the continuum approximation errors are of the same order of magnitude. Decreasing $H_1$ any further does not decrease the total approximation error. 
\item Choose a decreasing sequence of positive $t_*$-values starting from the one above and calculate the numerical approximations with the fixed $H_1$-value found in the previous step. If we choose $\beta$ as in \Eqref{eq:balancedbeta}, the numerical and continuum approximation errors should now be balanced in size.
\end{enumerate}
In some of our numerical experiments presented below we shall intentionally choose values of $H_1$ which are ``too large'' or ``too small''. We shall show that this allows us to study some \emph{intermediate} convergence behavior in agreement with the theory. We shall also often intentionally choose ``non-balanced'' $\beta$-values in order to fully validate the theoretical predictions.


\subsection{Derivation of the error estimate (\Theoremref{theorem:RK2})}
\label{sec:proof}

We now give a proof of \Theoremref{theorem:RK2}. 
We first note that under the hypothesis of the theorem,
\Eqref{eq:BoundSouceref} can be generalized to
\begin{equation}
  \label{eq:BoundSoucerefd1}
  \left|t^{-(\delta-l_1-|l_2|\mu_0)}(\partial_t^{l_1}\partial_u^{|l_2|}f)(t,y(t))\right|
  \le d+L s
\end{equation}
for any smooth  $y:(0,T]\to \R^n$ with
$\sup_{t\in(0,T]}|t^{-\mu_0} y(t)|\le s$ and for all non-negative integers
$l_1$ and $n$-dimensional multiindices $l_2$ with $l_1+|l_2|\le 3$.
The proof now splits
into the following natural steps.  
\newcounter{proofstep}
\setcounter{proofstep}{0}

\refstepcounter{proofstep}
\label{prf:step1}
\paragraph{Step~\theproofstep. Continuation criterion and local
  existence for the discrete approximations.}
We start by proving the following \emph{discrete continuation criterion}:
Given an arbitrary positive integer $N$, suppose the numerical
solution $(y_i)$ has been obtained using \Eqsref{eq:singlestepscheme} and
\eqref{eq:RK2} up to the $\widehat N$th time step for an arbitrary
$\widehat N\in [0,N)$, i.e., one has been able to calculate
$(y_i)_{i=0,\ldots,\widehat N}$, and that
$|t_{\widehat N}^{-\mu_0} y_{\widehat N}|\le s/2$. Then the next time step
given by \Eqsref{eq:singlestepscheme} and \eqref{eq:RK2} is
well-defined and finite, i.e., the sequence
$(y_i)_{i=0,\ldots,\widehat N+1}$ is finite. The proof for this statement is accomplished if,
 according to
\Eqref{eq:singlestepscheme}, \eqref{eq:RK2} and condition (i) in Section \ref{thm:SIVP}, it is possible to bound by $s$ the following quantity:  
\begin{equation*}
    \left|\left(1+\frac {\alpha_{\widehat N}}
    2\right)^{-\mu_0}t_{\widehat N}^{-\mu_0}\left[\left(1+\frac{A\alpha_{\widehat N}}2\right)y_{\widehat N}+\frac
    {\alpha_{\widehat N}} 2  f(t_{\widehat N},y_{\widehat
      N})\right]\right|. 
\end{equation*}
This expression is smaller or equal than
\begin{align}
  &\frac{|1+{A\alpha_{\widehat N}}/2|}
    {\left(1+{\alpha_{\widehat  N}}/2\right)^{\mu_0}}
    \left|t_{\widehat N}^{-\mu_0} y_{\widehat  N}\right|
    +\frac{t_{\widehat N}^{\delta-\mu_0}\alpha_{\widehat  N}/2}
    {\left(1+{\alpha_{\widehat N}}/2\right)^{\mu_0}}
    \left|t_{\widehat N}^{-\delta} f(t_{\widehat N},y_{\widehat N})\right|\notag\\
  \label{eq:intermedest3}
  \le&\frac{|1+{A\alpha_{\widehat N}}/2|}
       {\left(1+{\alpha_{\widehat  N}}/2\right)^{\mu_0}}\frac s2
       +\frac{\alpha_{\widehat N}/2}{\left(1+{\alpha_{\widehat N}}/2\right)^{\mu_0}}
       T^{\delta-\mu_0}\left(d+L \frac s2\right),
\end{align}
using \eqref{eq:BoundSouceref} in the last step. Since
$\alpha_{\widehat N}$ can be assumed to be arbitrarily small, this can be bounded by a smooth function in
$\alpha_{\widehat N}$ given that the matrix norm in the first
term can be bounded by the Frobenius norm which depends smoothly on
the entries of the matrix near the identity. This smooth function in
$\alpha_{\widehat N}$ therefore satisfies a Lipschitz property, which can
be exploited to bound the right hand side of \eqref{eq:intermedest3}
by a constant, which depends only on $A$, $\mu_0$, $\delta$, $T$, $d$,
$L$, $s$ and $H_1$ and $t_*^\beta$ (using
\Eqref{eq:alpharestrbounded}). In particular this constant can be made
arbitrarily small by choosing $H_1$ sufficiently small. Hence we can
indeed obtain the required bound $s$ if $H_1$
is sufficiently small.  This establishes the discrete continuation
criterion above.

A direct consequence of this continuation criterion is \emph{local
  existence} for the discrete evolution: Since
$|t_{0}^{-\mu_0} y_{0}|=0\le s/2$, the hypothesis of the continuation
criterion holds for $\widehat N=0$ and the numerical evolution can be
extended to at least the time step $\widehat N=1$.

\refstepcounter{proofstep}
\label{prf:step2}
\paragraph{Step~\theproofstep. Estimates for $K$.}
For this step of the proof it is convenient to consider $\alpha_i$ as
a smooth function $\alpha: (0,T]\to  0$ with
$\alpha(t_i)=\alpha_i$ for all $i=0,\ldots, N$ which satisfies the
bound \eqref{eq:alpharestrbounded} at all $t\in (0,T]$. We also pick
arbitrary smooth functions $z_1(t)$ and $z_2(t)$ defined on
$(0,T]$. Then, for every value of $t\in (0,T]$ for which
$|t^{-\mu_0} z_1(t)|\le s/2$ and $|t^{-\mu_0} z_2(t)|\le s/2$, we
conclude that
\[\left|t^{-\mu_0}\left(1+\frac {\alpha(t)} 2\right)^{-\mu_0}
    \widetilde z_1\left(\left(1+\frac {\alpha(t)}
        2\right)t\right)\right|\le s,\]
for all $t\in (0,T]$ where
\[\widetilde z_1\left(\left(1+\frac {\alpha(t)} 2\right)t\right)
  =\left(1+\frac{A\alpha(t)}2\right) z_1(t)+\frac {\alpha(t)} 2
  f(t,z_1(t)),\] using the same arguments as in
Step~\ref{prf:step1}. The corresponding function defined by $z_2$
satisfies the same bound. As a consequence of \Eqsref{eq:DefK},
\eqref{eq:RK2} and condition (i) in Section \ref{thm:SIVP}  the
following is therefore well-defined
\begin{align*}
&  (1+K(t,z_1(t),z_2(t); \alpha(t)))(z_1(t)-z_2(t))\\ 
   & = \frac{2+\alpha(t)(2A+1)+A^2 \alpha^2(t)}{2+\alpha(t)} (z_1(t)-z_2(t))
     +\frac {A\alpha^2(t)}{2+\alpha(t)}  (f(t,z_1(t))-f(t,z_2(t)))\\
     & \quad +
       \frac{2\alpha(t)} {2+\alpha(t)} \Biggl[f\left(\left(1+\frac {\alpha(t)} 2\right)t,
       \left(1+\frac{A\alpha(t)}2\right) z_1(t)+\frac {\alpha(t)}2  f(t,z_1(t))\right)\\      
     &\qquad\qquad \qquad \quad -
       f\left(\left(1+\frac {\alpha(t)} 2\right)t,
       \left(1+\frac{A\alpha(t)}2\right) z_2(t)+\frac {\alpha(t)} 2  f(t,z_2(t))\right)\Biggr].
\end{align*}
Given
\Eqref{eq:LipschitzSourcePW} and the bounds on $\widetilde z_1$ and
$\widetilde z_2$ above, it follows that
\begin{align*}
  &\left(1+\frac {\alpha(t)} 2\right)^{-\delta}t^{-\delta} 
  \left|f\left(\left(1+\frac {\alpha(t)} 2\right)t,
        \widetilde z_1 \left(\left(1+\frac {\alpha(t)} 2\right)t\right)\right)      
    -
      f\left(\left(1+\frac {\alpha(t)} 2\right)t,
        \widetilde z_2 \left(\left(1+\frac {\alpha(t)} 2\right)t\right)\right)
\right|\\
&\le L \left(1+\frac {\alpha(t)} 2\right)^{-\mu_0}t^{-\mu_0} 
  \left|\widetilde z_1 \left(\left(1+\frac {\alpha(t)} 2\right)t\right)     
    -      
        \widetilde z_2 \left(\left(1+\frac {\alpha(t)} 2\right)t\right)
\right|\\
&\le L \left(1+{\alpha(t)/2}\right)^{-\mu_0}\left( (1+{A\alpha(t)}/2)
  t^{-\mu_0}|z_1(t)-z_2(t)|         
+\frac{\alpha(t)}2t^{\delta-\mu_0}t^{-\delta}|f(t,z_1(t))-f(t,z_2(t))|\right)
\\
&\le C \left( \frac{|1+{A\alpha(t)/2|}}{\left(1+{\alpha(t)/2}\right)^{\mu_0}}    
+\frac{\alpha(t)/2}{\left(1+{\alpha(t)/2}\right)^{\mu_0}} t^{\delta-\mu_0}\right) t^{-\mu_0}|z_1(t)-z_2(t)|,
\end{align*}
where $C$ is a constant which depends only on $L$. We conclude that
\begin{align*}
&\left|(1+K(t,z_1,z_2; \alpha))\right|
\\
&    \le
\left|\frac{2+\alpha(2A+1)+A^2 \alpha^2}{2+\alpha}\right|
     +\frac {\alpha^2}{2+\alpha}t^{\delta-\mu}|A|
+
       C\frac{2\alpha} {2+\alpha} \left(1+\frac {\alpha} 2\right)^{\delta-\mu_0}t^{\delta-\mu_0}
\left( {\left|1+A \frac\alpha2\right|}    
+\frac{\alpha}{2} t^{\delta-\mu_0}\right)\\
&\le
1+\alpha \left(|A|+\frac{2 \left(1+\frac {\alpha} 2\right)^{\delta-\mu_0}T^{\delta-\mu_0}} {2+\alpha} +\alpha\frac{|A(A-1)| +T^{\delta-\mu}|A|+
       C\left(1+\frac {\alpha} 2\right)^{\delta-\mu_0}T^{\delta-\mu_0}
\left(|A|  
+T^{\delta-\mu_0}\right)}{2+\alpha}\right).
\end{align*}
The arguments above therefore yield 
$
  \left|(1+K(t,y_1(t),y_2(t); \alpha(t)))\right|
  \le 1+k'\alpha(t),
$  
where $k'>|A|$ and, by a judicious choice of $T$ and $H_1$, we can
make $k'$ arbitrarily close to $|A|$.

\refstepcounter{proofstep}
\label{prf:step3}
\paragraph{Step~\theproofstep. Global existence for the discrete
  approximations.}
Being equipped with the local existence and continuation results from
Step~\ref{prf:step1} and the estimates for $K$ in Step~\ref{prf:step2}, we can
now  tackle the global existence problem for the discrete
evolution. By this we mean the existence of a time interval
$(0,\widehat T]$ with $\widehat T\le T$ such that discrete evolutions exist
for all $i=0,\ldots,N$ and are bounded in a way which is independent
of the choice of $t_*$ and of $N$.  By the local existence result we
know that there is a \emph{maximal} integer $\widehat N$ in $[1,N]$ such
that the sequence $(y_i)_{i=0,\ldots,\widehat N}$ is well-defined by
\Eqsref{eq:singlestepscheme} and \eqref{eq:RK2} and is finite. Since
$\widehat N$ is assumed to be maximal, the continuation criterion in
Step~\ref{prf:step1} implies that either $\widehat N=N$, or, if
$\widehat N<N$, then $|t_{\widehat N}^{-\mu_0} y_{\widehat N}|> s/2$. We establish
now by a judicious choice of $T$ and $H_1$ that the latter is
impossible. In order to establish the corresponding contradiction let
us suppose that the latter holds.  Observe that
\Eqsref{eq:singlestepscheme} together with \Eqsref{eq:DefK} and
\eqref{eq:defweight} implies that (using the same convention for $\Pi$
as for \eqref{eq:errorestimate5555})
  \[w_i y_i=\sum_{l=0}^{i-1} \omega_l w_{l} \Phi_l
    \prod_{j=l+1}^{i-1} \left(1+K^{(0)}_j\right) \omega_j,\quad w_0y_0=0,\]
  for all $i=1,\ldots, \widehat N$, where
  \begin{equation}
    \label{eq:shorthand2}
    \Phi_l=\Phi(t_l,0;\alpha_l),\quad K^{(0)}_j=K(t_j,y_j,0;\alpha_j).
  \end{equation}
  Given the initial condition we know that there is a maximal integer
  $\check N$ in $[0,\widehat N)$ such that
  $\sup_{i=0,\ldots,\check N}|t_{i}^{-\mu_0} y_{i}|\le s/4$. The
  estimate for $K$ established in Step~\ref{prf:step2} yields the
  existence of  $k'>|A|$ (which we can assume to be arbitrarily
  close to $|A|$ under the conditions here) such that
$|1+K^{(0)}_j|\le 1+k'\alpha_j$
  for all $j=0, \ldots, \check N$. Given 
  \begin{equation}
    \label{eq:weightexponent}
    \omega_i=(1+\alpha_i)^{-\lambda},
  \end{equation}
  which is compatible with the choice $w_i=t_i^{-\lambda}$ and
  \Eqref{eq:defweight}, it is elementary to check that 
  \[|1+K^{(0)}_j|\omega_j\le 1\]
  for any $\lambda>|A|$  provided  $T$ and $H_1$ are
  sufficiently small. As a consequence
  \begin{equation}
    \label{eq:bootstrapgrowth1}
    |w_i y_i|\le \sum_{l=0}^{i-1} |\omega_l w_{l} \Phi_l|,\quad |w_0
    y_0|=0,
  \end{equation}
  for all $i=1,\ldots, \check N$. \Eqsref{eq:weightexponent},
  \eqref{eq:shorthand2}, \eqref{eq:RK2} and \eqref{eq:defd} yield, for
  any $\epsilon>0$ and all $l=0,\ldots,\check N$,
  \begin{align*}
    \frac{t_l^{-\epsilon}|\omega_l w_{l}\Phi_l|}{(1+\alpha_l)^\varepsilon-1}
    \le &\frac{2|A| t_l^{\delta-\lambda-\epsilon}\alpha_l}{\epsilon}\frac{(1+\alpha_l/2)^{-1}}{(1+\alpha_l)^{\lambda}}
    \frac {\epsilon\alpha_l}{(1+\alpha_l)^\varepsilon-1} 
      d\\
      &+\frac{t_l^{\delta-\lambda -\epsilon}}{\epsilon}\frac{(1+\alpha_l/2)^{\delta-1}}{(1+\alpha_l)^{\lambda}}
    \frac {\epsilon\alpha_l}{(1+\alpha_l)^\varepsilon-1}
      \underbrace{\left|t_l^{-\delta}\left(1+\frac {\alpha_l} 2\right)^{-\delta}f\left(\left(1+\frac {\alpha_l} 2\right)t_l,
        \frac {\alpha_l} 2  f(t_l,0)\right)\right|}_{\le D},
  \end{align*}
where $D$ is a constant which is independent of $t_l$ and $\alpha_l$
in particular as a consequence of the fact that we establish the bound
\[\left|t_l^{-\mu_0}\left(1+\frac {\alpha_l} 2\right)^{-\mu_0}\frac
    {\alpha_l} 2  f(t_l,0)\right|\le s\]
for any $t_l$ and $\alpha_l$ provided $H_1$ and $T$ are sufficiently small. 
The same arguments as before yield that the quantity
$\frac{t_l^{-\epsilon}|\omega_l
  w_{l}\Phi_l|}{(1+\alpha_l)^\varepsilon-1}$ can be estimated by
 a constant $C>0$ which can be chosen
arbitrarily small, provided $|A|<\lambda<\delta$,
$0<\epsilon<\delta-\lambda$, and, $T$ and $H_1$ are
sufficiently small. This together with \Eqsref{eq:bootstrapgrowth1}
and \eqref{eq:integralestimate} and the continuation criterion therefore yields 
$
  \sup_{i=0,\ldots\check N+1}|w_i y_i|\le C T^\epsilon.
$

Hence, we have found that given $\lambda$ and $\epsilon$ in the
ranges above we can always find sufficiently small $T$ and $H_1$ such
that $\sup_{i=0,\ldots\check N+1}|w_i y_i|\le s/4$ independently of the choice of
$t_*\in (0,T]$. This contradicts the assumption that $\check N$ is the
\emph{maximal} number in $[0,\widehat N)$ with the property
$\sup_{i=0,\ldots\check N}|w_i y_i|\le s/4$. Such a maximal integer
therefore does not exist. This implies that
$\sup_{i=0,\ldots\widehat N}|w_i y_i|\le s/4$, and, this contradicts the
assumption that $|w_{\widehat N} y_{\widehat N}|>s/2$. The integer $\widehat N$
can therefore \emph{not be maximal and satisfy $\widehat N<N$}. It follows that
$\widehat N=N$.

\refstepcounter{proofstep}
\paragraph{Step~\theproofstep. Estimates for the truncation error and
  the numerical approximation error.}
Given now that the discrete evolutions exist independently of the
choice of $t_*$ and $N$ on a uniform time interval we can now finally
proceed with estimating the numerical approximation error, see
\Eqref{eq:errorestimate39842}. Given the estimate for $K$ before, the
main remaining task is to estimate the truncation error $T_l$, see
\Eqsref{eq:truncationerror} and \eqref{eq:notation}, for RK2 given by \Eqref{eq:RK2}. The well-known fact that that RK2
is a scheme of order $2$ can be checked directly by lengthy
calculations and can be expressed in the following form
\[\Theta(t,y;\alpha)=\alpha^3\int_0^1\int_0^1\int_0^1 x_1^2 x_2\,
  T_{\alpha,\alpha,\alpha} (t,y;\alpha\, x_1 x_2 x_3)\, dx_3 dx_2
  dx_1\] where $T_{\alpha,\alpha,\alpha}$ is the third
$\alpha$-derivative of the function $T$. Pick now an arbitrary $t_*$
on our time interval. Let $v: (0,T]\to \R$ be the function
(which had been labelled $v_{t_*}$ in \Theoremref{thm:approx}) where
we suppose (by shrinking $T$ if necessary) that $\tildeT=T$.  Let
$\alpha(t)$ be defined in the same way as in Step~\ref{prf:step2}. By
further lengthy calculations using \Eqsref{eq:BoundSoucerefd1},
\eqref{eq:CPsolest} and \eqref{eq:LipschitzSourcePWDer}, we can show that
there are exponents $\kappa_1$ and $\kappa_2$, which only depend on
$\delta$ and $\mu_0$ (more details below), and a constant $C>0$, which
only depends on $d$, $L$, $s$ and $\delta$ and $A$, such that
\[|T_{\alpha,\alpha,\alpha} (t,v(t);\alpha(t)\, x)|\le C
  t^{\delta}(1+x\alpha/2)^{-\kappa_1} (1+x\alpha)^{\kappa_2},\]
where we write $x=x_1 x_2 x_3\in [0,1]$. Given this bound for $x$ and
the bound for $\alpha$ given by \Eqsref{eq:alpharestr} and
\eqref{eq:boundednessalpha}, we can bound 
\[(1+x\alpha/2)^{-\kappa_1} (1+x\alpha)^{\kappa_2}\]
by a constant, which only depends on $d$, $L$, $s$, $\delta$,
$A$, $\mu_0$ and $H_1$ and which can be chosen arbitrarily close to $1$
by choosing $H_1$ to be sufficiently small. This means that 
$
|T (t,v(t);\alpha(t))|\le C
  t^{\delta} \alpha^{3}(t)
  $
for a constant $C>0$ with the same dependencies. 

Regarding the numerical approximation error
\Eqref{eq:errorestimate39842}, we first notice, similar to
Step~\ref{prf:step3}, that
$\prod_{j=l+1}^{N-1} \left|(1+K_j) \omega_j\right|\le 1$ given
\eqref{eq:weightexponent} and $\lambda>|A|$ as a consequence of
Step~\ref{prf:step2}. \Eqref{eq:errorestimate39842} therefore reduces
to
\begin{equation*}
  \sup_{i=0,\ldots,N}|w_{i}(v(t_{i})-y_{i})|\le \sum_{l=0}^{N-1} |\omega_l w_{l} T_l|
\end{equation*}
with $T_l=T (t_l,v(t_l);\alpha_l)$. With a view to applying
\Eqref{eq:integralestimate} for some $\epsilon\not= 0$ we now estimate
\[|T(t,v(t);\alpha(t))|t^{-\lambda}(1+\alpha(t))^{-\lambda}\frac{1}{(1+\alpha(t))^\epsilon-1}
  \le C t^{\delta-\lambda} \alpha^{2}(t)
  (1+\alpha(t))^{-\lambda}\frac{\alpha(t)}{(1+\alpha(t))^\epsilon-1}.\]
The same arguments as before (observe that $x/((1+x)^\epsilon-1)$ is a
smooth function near $x=0$), it follows under the same assumptions as
before for any $\epsilon\not=0$ that
\[(1+\alpha(t))^{-\lambda}\frac{\alpha(t)}{(1+\alpha(t))^\epsilon-1}\]
can be bounded by a constant arbitrarily close to $1/|\epsilon|$
whenever $H_1$ is sufficiently small. Then
\[|T_l\omega_l w_l| \le \frac C\epsilon H_1^2
  t_*^{2(\beta+1-\eta)}t_l^{\delta-\lambda+2(\eta-1)}
  ((1+\alpha_l)^\epsilon-1).\] Setting
$\epsilon=\delta-\lambda+2(\eta-1)$ and assuming that this is not
zero, \Eqref{eq:integralestimate} yields that
\begin{equation*}
  \sup_{i=0,\ldots,N}|w_{i}(v(t_{i})-y_{i})|\le \frac C{|\delta-\lambda+2(\eta-1)|} H_1^2
  t_*^{2(\beta+1-\eta)} |T^{\delta-\lambda+2(\eta-1)}-t_*^{\delta-\lambda+2(\eta-1)}|,
\end{equation*}
where $C$ is a constant, which only depends on $d$, $L$, $s$, $\delta$,
$A$, $\mu_0$ and $H_1$ and which can be chosen arbitrarily close to $1$
by choosing $H_1$ to be sufficiently small. This implies
\Eqsref{eq:mainerrorestimate00}.


\section{Numerical investigation of Fuchsian equations} 
\label{sec:Numtestproblem}

\subsection{A model problem}
\label{sec:ODEtestproblem}

\paragraph{An explicit formula for a Fuchsian model problem}

We begin with a relatively simple (but non-trivial) problem which is going to numerically confirm our theoretical
analysis in \Sectionref{sec:mainanalyticalresult}. The test problem
is defined by \Eqref{eq:FuchsianODEgen} with $n=1$ and, for an arbitrary constant $p>0$, 
\begin{equation}
  \label{eq:modelprobldata}
  A=0,\qquad f(t,y)=p(y^2+t^{2p}),
\end{equation}
hence 
\begin{equation}
  \label{eq:FuchsianODEgen-test}
  t\partial_t u = p \, (u^2+t^{2p}),\qquad t>0. 
\end{equation}
We could in principle assume here that $p=1$ without loss of generality. This is so since, given some $p>0$,
we can always apply a transformation of the form \Eqref{eq:transf01} and produce the same type of the equation but 
with $p=1$. However, since the RK2-scheme is not invariant under such transformations, it is useful to keep $p$
as an arbitrary positive parameter.

First observe that \Eqref{eq:regularcond1} is violated for any
$\delta>0$.  The singular initial value problem can therefore not be
transformed to a regular Cauchy problem via \Eqsref{eq:transf01} and \eqref{eq:RegCauchyProblTrafo}. In fact, this Cauchy problem would have
infinitely many different solutions since all solutions of this
equations tend to zero at $t=0$. This can be seen by taking the limit at $t=0$ of the  
general solution  
\begin{equation}
  \label{eq:modelproblexactGen} 
  v(t)=t^p\frac{ c \text{Y}_1(t^p )+\text{J}_1(t^p)} {c\text{Y}_0(t^p)+\text{J}_0(t^p )}
\end{equation}
of \Eqref{eq:modelprobldata}
determined by an arbitrary constant $c\in\R$, where
$\text{J}_n$ and $\text{Y}_n$ are \emph{Bessel functions} of first and
second kind, respectively. We show that \Eqref{eq:modelproblexactGen} is the general solution by a straightforward
calculation, where we first introduce $\tau=t^p$ as a new time
variable (which transforms the equation given by a general $p>0$ to that
with $p=1$). Second we write $v(\tau)=\tau Z_1(\tau)/Z_0(\tau)$ where
$Z_k(\tau)=\text{J}_k(\tau)+c \text{Y}_k(\tau)$, and finally we check that
\[\tau\partial_\tau\left(\tau \frac{Z_1}{Z_0}\right)-\tau^2 \frac{Z_1^2+Z_0^2}{Z_0^2}=0\]
by using the standard relations for Bessel functions (for arbitrary integers $k$)
\[\frac{2kZ_k(\tau)}{\tau}=Z_{k-1}(\tau)+Z_{k+1}(\tau),\quad
  2\frac{dZ_k(\tau)}{d\tau}=Z_{k-1}(\tau)-Z_{k+1}(\tau).
  \]


The setup in Section \ref{thm:SIVP} applies with $\delta=2p$ and $\mu_0=p$. The
unique solution $u=u(t)$ to the singular initial value
problem turns out to be obtained by setting $c=0$ in \Eqref{eq:modelproblexactGen}, that is, 
\begin{equation}
  \label{eq:modelproblexactSP} 
  u(t)=t^p\frac{\text{J}_1(t^p)}
  {\text{J}_0(t^p )}.
\end{equation}
All other solutions other than $u(t)$ would have a non-trivial $1/\log t$ factor and therefore a log behavior at
$t=0$. 

\begin{figure}[ht]
  \centering
  \includegraphics[width=0.6\textwidth]{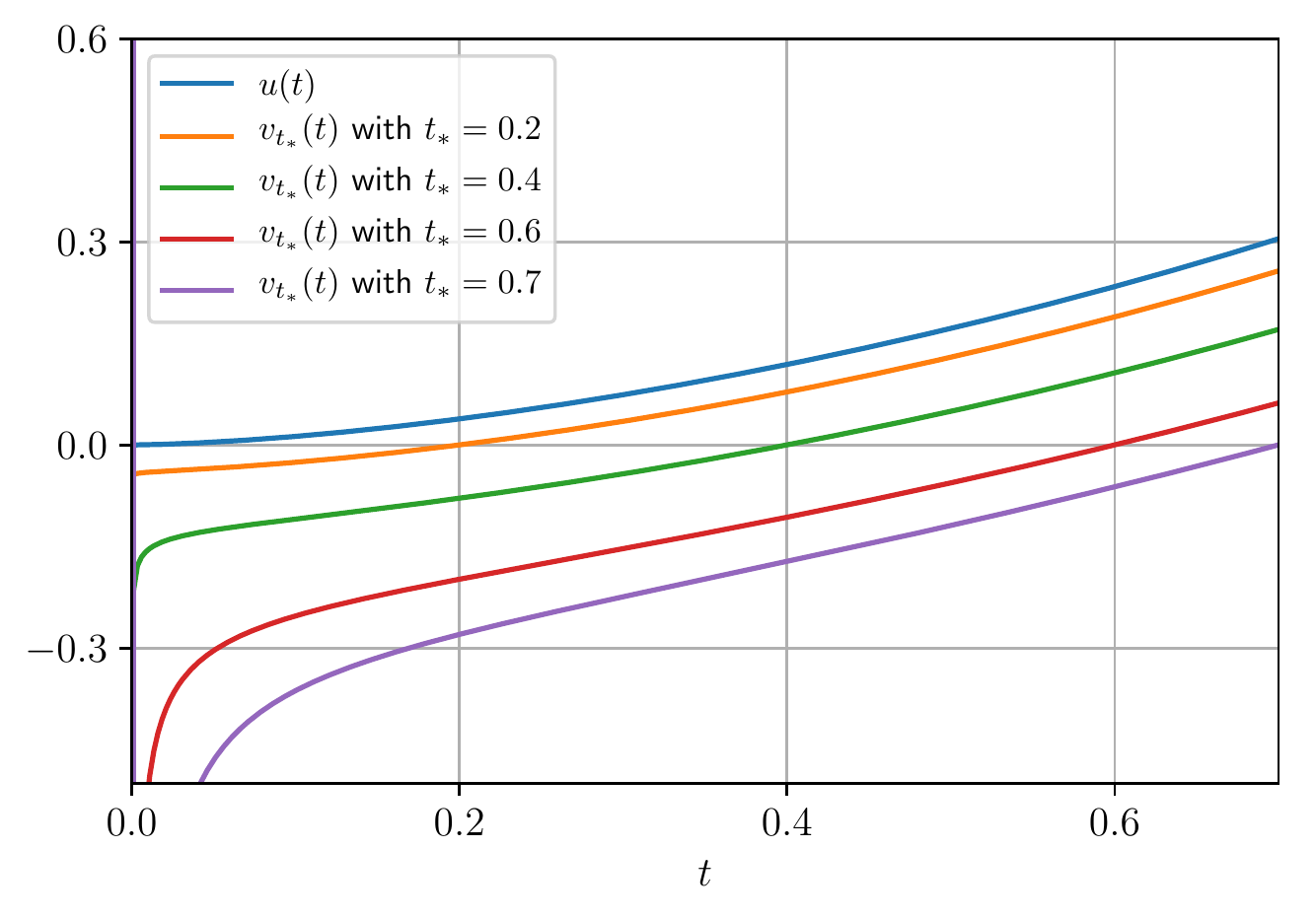}
  \caption{Exact solutions $v_{t_*}$ of the test problem
    \Eqref{eq:modelprobldata} given by \Eqsref{eq:modelproblexactSP} and \eqref{eq:modelproblexactCP} for $p=0.8$.}
  \label{fig:graphs}
\end{figure}

According to \Theoremref{thm:approx}, the solution $u(t)$ in \Eqref{eq:modelproblexactSP} can be
approximated by solutions $v_{t_*}(t)$ of the regular Cauchy problem
problem with $v_{t_*}(t_*)=0$. From \Eqref{eq:modelproblexactGen} we derive that any such solution takes the explicit form
\begin{equation}
  \label{eq:modelproblexactCP}
  v_{t_*}(t)=t^p \frac{\text{J}_1\left(t_*^p\right)
    \text{Y}_1\left(t^p\right)-\text{J}_1\left(t^p\right)
    \text{Y}_1\left(t_*^p\right)}{\text{J}_1\left(t_*^p\right)
    \text{Y}_0\left(t^p\right)-\text{J}_0\left(t^p\right)
    \text{Y}_1\left(t_*^p\right)},
\end{equation}
see \Figref{fig:graphs}. Observe that all graphs in this figure
approach zero at $t=0$. It is a consequence of the $1/\log t$-behavior
for most of the graphs that this approach is very slow and therefore not
properly resolved in the figure.

\paragraph{Applying transformations to the model problem}

As an instructive side step, it also makes sense to
apply the transformation
\begin{equation}
  \label{eq:modelproblemtrafo}
  z(t)=1/y(t)
\end{equation}
to \Eqref{eq:modelprobldata} which takes the form
\[t \partial_t z(t)=-p(1+ t^{2p} z^2(t)).\] 
We find that the setup in Section \ref{thm:SIVP}  does not apply to this equation directly. However if we
set
\begin{equation}
  \label{eq:modelsplit}
  z(t)=a t^{-2p}+w(t),
\end{equation}
the equation implies that either $a=0$ or $a=2$, and that
\[t\partial_t w(t)=-p(1+2 a w(t)+t^{2p} w(t)^2).\] 

\bei 

\item 
For $a=2$, 
Section \ref{thm:SIVP} applies with $A=-4p$ and $-3p<\mu_0<\delta$. It turns out that the uniquely determined solution $w(t)$ agrees with \Eqref{eq:modelproblexactSP}
once $w$ has been transformed back using \Eqsref{eq:modelsplit} and \eqref{eq:modelproblemtrafo}.
The case $a=2$ does therefore not yield anything new.

\item 
Consider now the
case $a=0$. Again, Section \ref{thm:SIVP} does not apply directly. But
if we set
\[z(t)=a t^{-2p}-p\log t+z_*+w(t)\] for a (so far arbitrary) constant
$z_*$, we get
\[t\partial_t w(t)=-p\,t^{2p} (z_*-p\log t+w(t))^2.\]
Section \ref{thm:SIVP} now applies with $A=0$, $\delta<2p$ and
$0<\mu_0<\delta$. Interestingly, in contrast to the case $a=2$ above,
it is now possible to transform this problem to a {\sl regular} Cauchy problem using a transformation of the form \Eqsref{eq:transf01} and \eqref{eq:RegCauchyProblTrafo}, namely 
\be
\partial_{\widetilde t} {\widetilde u}({\widetilde t})=F({\widetilde t},{\widetilde u}({\widetilde t})),\quad {\widetilde u}(0)=0,
\ee
for 
\[F({\widetilde t},y)=-\frac p\delta{\widetilde t}^{(2p-\delta)/\delta}\left(z_*-\frac
    p{\delta}\log {\widetilde t}+y\right)^2,\] where
${\widetilde u}({\widetilde t})=w({\widetilde t}^{1/\delta})$. Given that the regularity of this
nonlinear function is low at ${\widetilde t}=0$, in particular if $\delta$ is
chosen to be close to $2p$ and if $p$ is small, it would be
interesting to compare the numerical errors when this equation is
solved with a standard regular Cauchy problem method as opposed to the method
for singular initial value problems discussed in this paper. (We do not pursue this issue here.)
\eei

\begin{figure}[t!]
  \hfill
  \begin{minipage}[t]{.49\textwidth}
    \centering
    \vspace{0pt}
    \includegraphics[height=5.3cm]
    {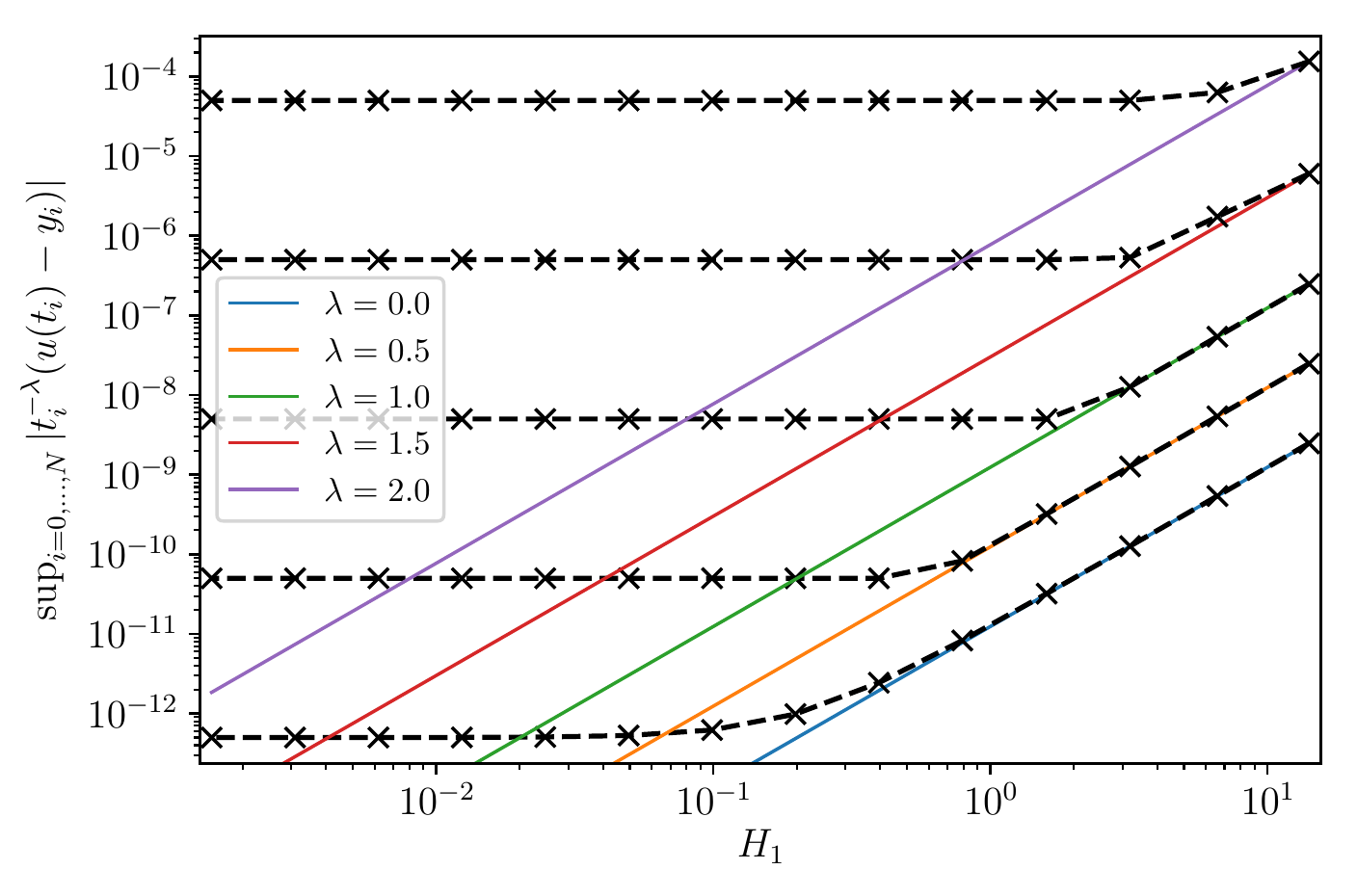}           
  \end{minipage}%
  \hspace*{\fill}

  \begin{minipage}[t]{.49\textwidth}
    \centering    
    \includegraphics[height=5.3cm]
    {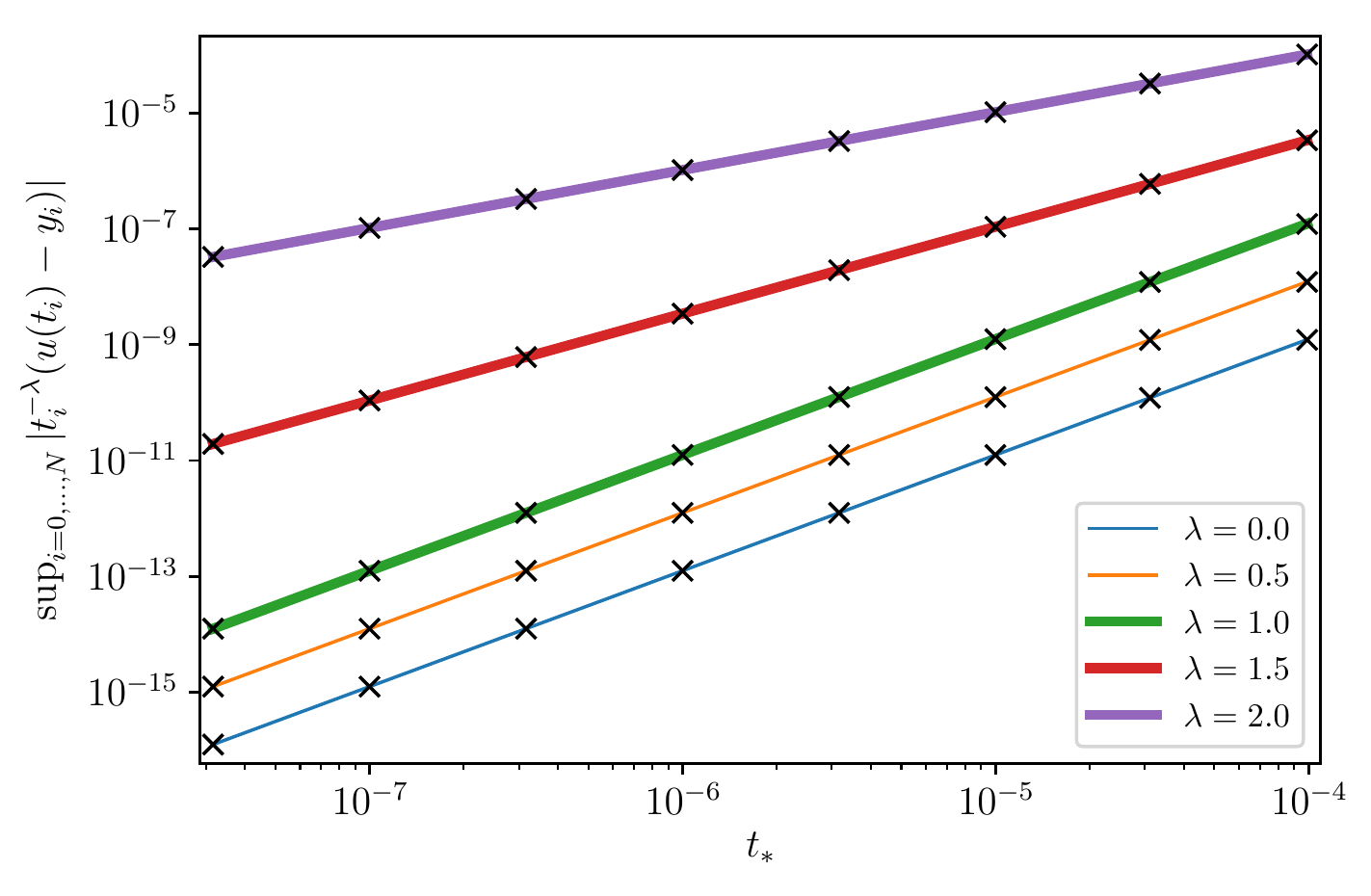}         
  \end{minipage}
  \begin{minipage}[t]{.49\textwidth}
    \centering 
    \includegraphics[height=5.3cm]
    {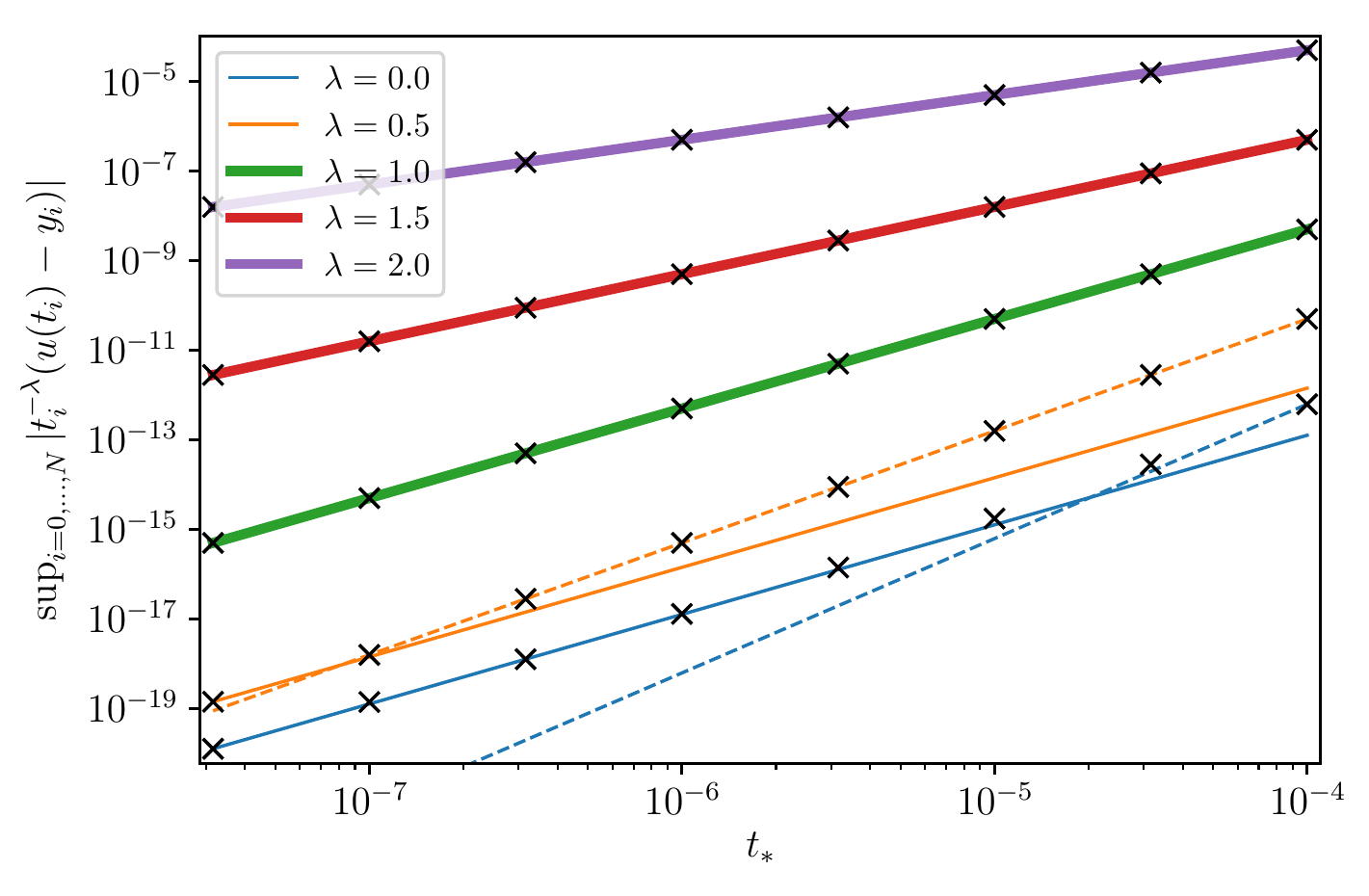}  
  \end{minipage}
\caption{Singular initial value problem \Eqref{eq:modelprobldata} for $p=1.5$: In the first plot, we  identify the ``optimal'' values for $H_1$ for various choices of $\lambda$, and for fixed values $t_*=10^{-4}$, $T=0.01$ and $\eta=\beta=0$. The continuous reference curves show the $O(H_1^2)$ behavior expected from \Theoremref{theorem:RK2} in regimes where the numerical approximation error is dominant. In the second plot, we show convergence for fixed $H_1=10$, $T=0.01$ and $\eta=\beta=0$ and a sequence of $t_*$-values  of the numerical solutions to the solution $u$ of the singular initial value problem given by
      \Eqref{eq:modelproblexactSP}. The total approximation errors are marked again by $\times$ while the reference curves show the expected
      $O(t_*^\sigma)$ behavior with $\sigma$ given by \Eqref{eq:sigmaest}. The third figure shows the same for $H_1=0.1$ (see
      the text for additional details). In all these plots, the total approximation errors are calculated with respect to the exact solution $u$ given by      \Eqref{eq:modelproblexactSP}. }  
\label{fig:numN1}
\end{figure}

\subsection{Numerical validation of the theory}
\label{numerical-analysis-1}

We still rely on the model \Eqref{eq:modelprobldata} in order to check our theoretical conclusions in \Sectionref{sec:mainanalyticalresult}.
In \Figref{fig:numN1} we show the numerical results for \(p=1.5\) and \(\beta=\eta=0\) based on the algorithm outlined at the end of \Sectionref{sec:discparam}.
The first plot shows Steps 1, 2, and 3, while the second and third plots show Step 4 for two different (fixed) values of $H_1$, respectively. In the first plot we clearly see that the ``optimal'' value for $H_1$ depends on the choice of $\lambda$. 
Since $\delta=2p=3$ and \(\beta=\eta=0\), the theoretical results
in \Sectionref{sec:mainanalyticalresult} reveal that \(\sigma_{num}<\sigma_{cont}\) if \(\lambda<1\), and,
\(\sigma_{num}=\sigma_{cont}\) --- the {\sl asymptotically balanced case} 
--- if \(\lambda\ge 1\).  All the numerical cases in \Figref{fig:numN1} are therefore either numerical error dominated ($\lambda<1$) or asymptotically balanced ($\lambda\ge 1$). In all of the following convergence plots, in particular therefore the second and third plots in \Figref{fig:numN1}, 
we indicate 
\bei 

\item asymptotically balanced cases by thick reference curves,

\item  numerical approximation error dominated cases by thin continuous reference curves 

\item and continuum error dominated cases by thin dashed reference curves. 

\eei 
Importantly, all these plots agree with the theoretical predictions very well. It is interesting to notice in the third plot of \Figref{fig:numN1}, where the numerical approximation error is small by virtue of the ``small'' value $H_1=0.1$, that the total approximation error is dominated by the continuum approximation error for intermediate values of $t_*$ for $\lambda=0$ and $\lambda=0.5$, before the theoretically predicted asymptotic numerical approximation error decay rate takes over for all sufficiently small values of $t_*$.

\begin{figure}[t!]
  \hfill
  \begin{minipage}[t]{.49\textwidth}
    \centering
    \vspace{0pt}
    \includegraphics[height=5.3cm]%
    {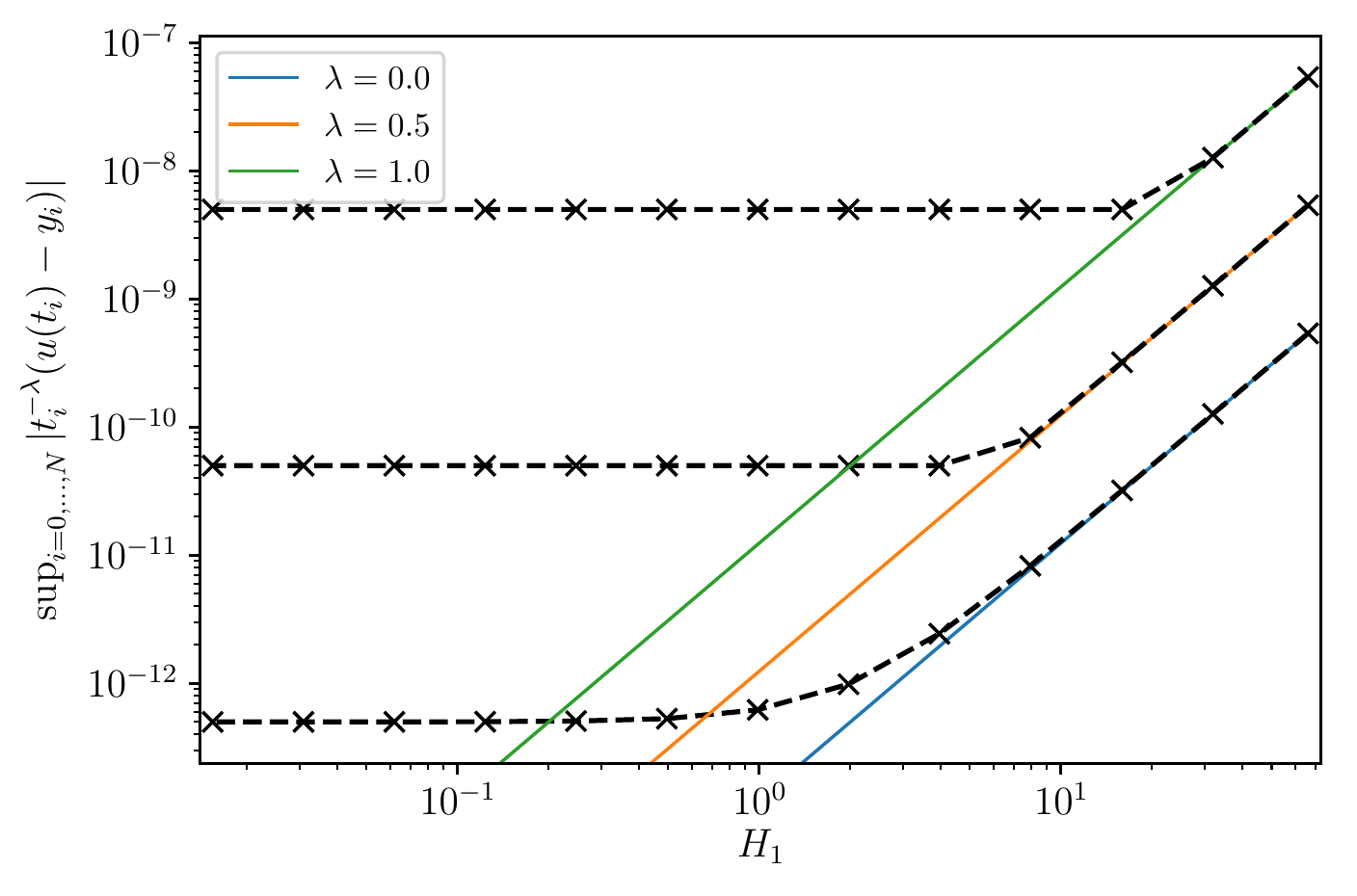}           
  \end{minipage}%
  \hspace*{\fill}

  \begin{minipage}[t]{.49\textwidth}
    \centering   
    \includegraphics[height=5.3cm]%
    {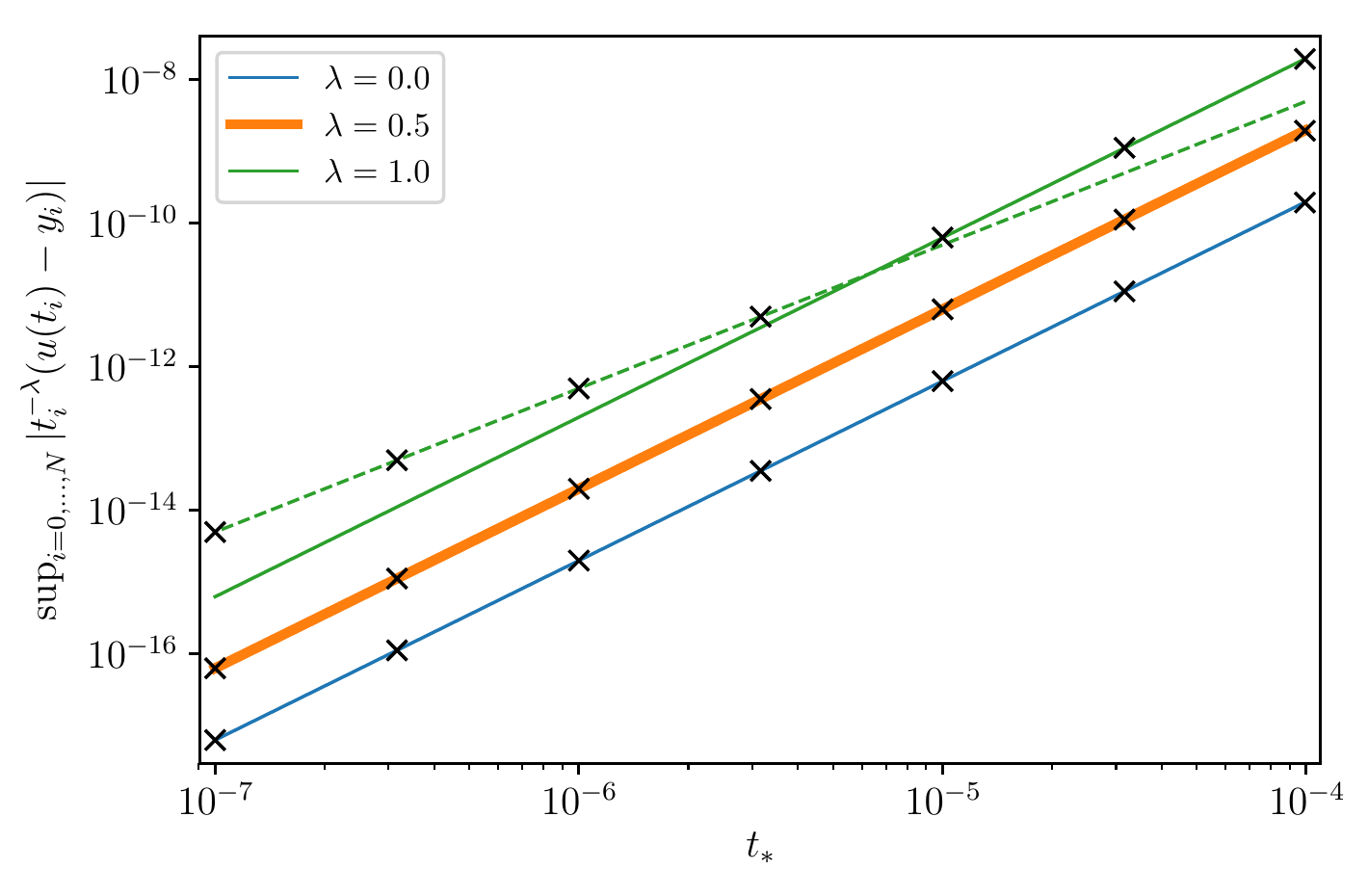}         
  \end{minipage}%
  \begin{minipage}[t]{.49\textwidth}
    \centering
    \includegraphics[height=5.3cm]%
    {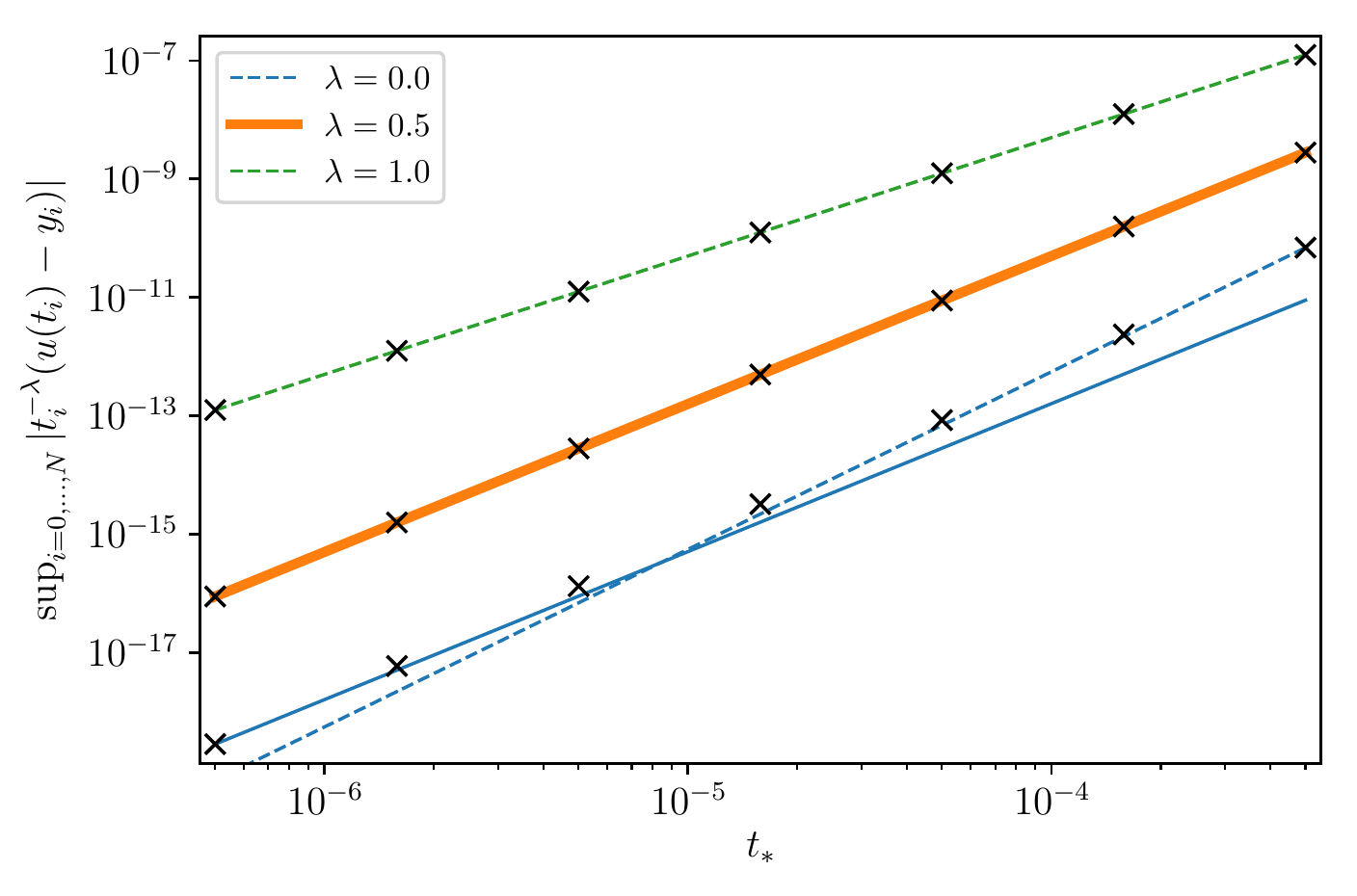}  
  \end{minipage}
  \caption{The same as \Figref{fig:numN1} but with  $\beta=0.25$. In the second and third plots we use $H_1=40$ and $H_1=1.0$, respectively.}
\label{fig:numN4}
\end{figure}

\begin{figure}[t!]
  \hfill  
  \begin{minipage}[t]{.49\textwidth}
    \centering   
    \includegraphics[height=5.3cm]%
    {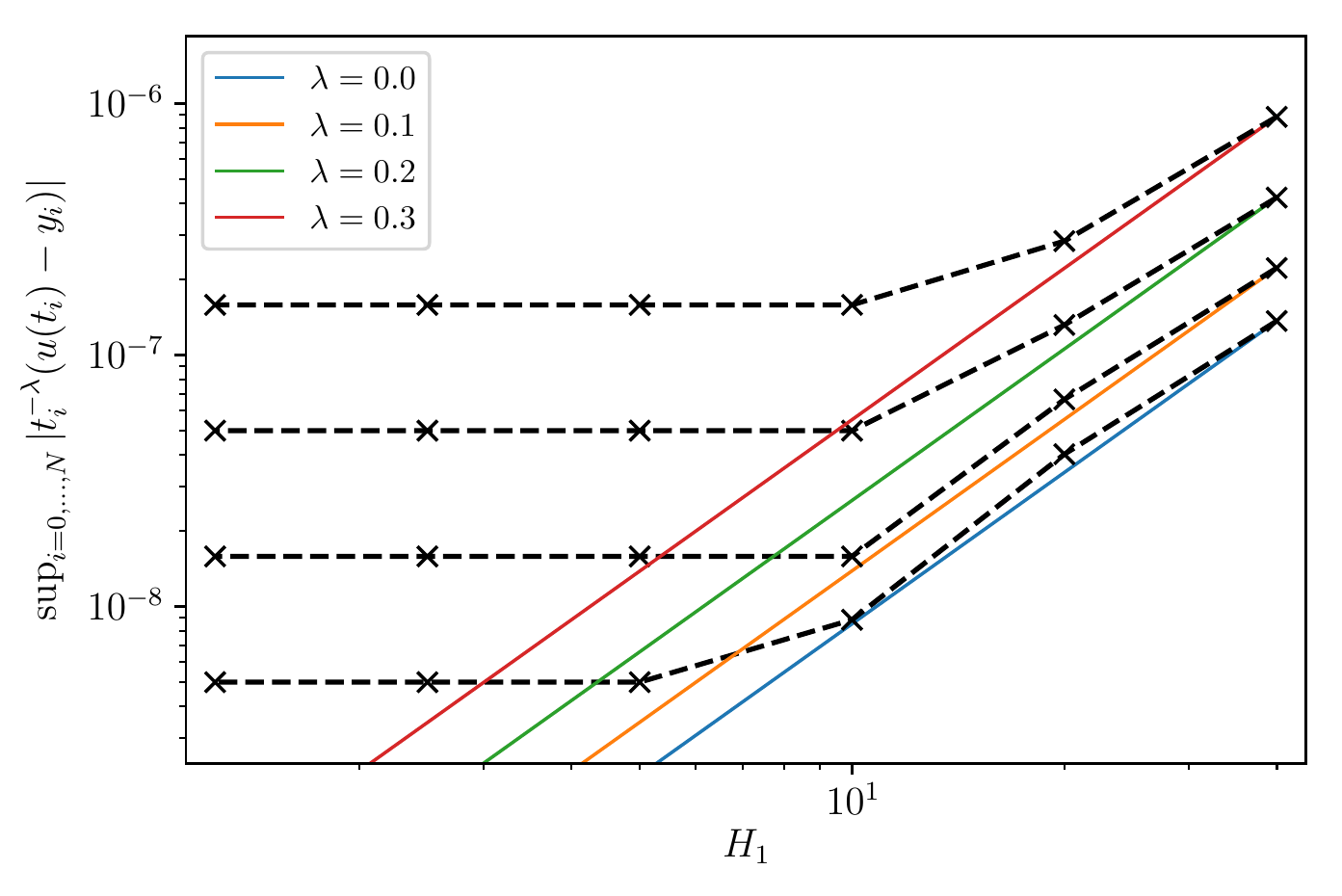}         
  \end{minipage}%
  \begin{minipage}[t]{.49\textwidth}
    \centering
    \includegraphics[height=5.3cm]%
    {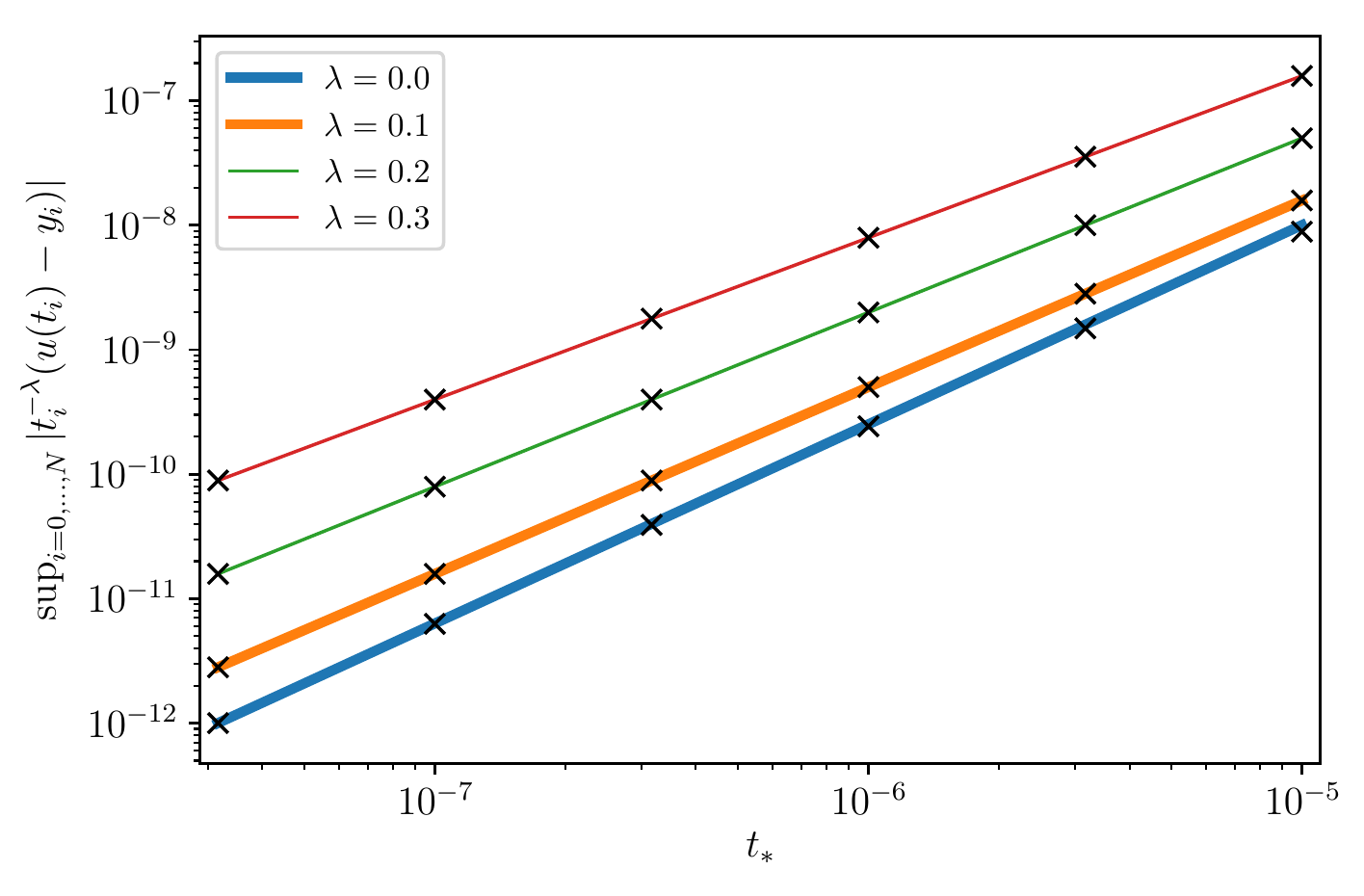}  
  \end{minipage}
  \caption{The same as \Figref{fig:numN1} but with $p=0.8$. In the first plot, we use $t_*=10^{-5}$. In the second plot we use $H_1=10$.}
\label{fig:numN7}
\end{figure}

Next we repeat the same numerical experiment for $p=1.5$ and $\beta=0.25$; see
\Figref{fig:numN4}. And then we pick up \(p=0.8\) and
\(\beta=\eta=0\); see \Figref{fig:numN7}. Our results are in full agreement with the theoretical analysis, and 
the last case also confirms that the sizes of the errors are larger the smaller $p$, and therefore $\delta$, is.

\subsection{Beyond the theoretical results: the case \texorpdfstring{$\beta\in (-1,0)$}{-1<beta<0}} 
\label{sec:numnegbeta}

The
hypothesis of \Theoremref{theorem:RK2} is not compatible with
$\beta<0$ since this would allow the sequence $\alpha_i$ to be
unbounded as a consequence of \Eqsref{eq:alpharestr} and
  \eqref{eq:boundednessalpha}. The discussion at the end
  of \Sectionref{sec:mainanalyticalresult} however indicates that the case $\beta<0$ might nevertheless be useful and potentially increase the numerical efficiency. 
Despite the fact that we are not able to prove an estimate for the numerical approximation error if $\beta<0$, it turns out that we manage to explore this case numerically.

Our numerical results indicate that the estimate \eqref{eq:mainerrorestimate} for the numerical approximation error holds also if
$\beta<0$, but with
\begin{equation}
  \label{eq:mainerrorestimateexpbetaneg}
  \sigma_{num}=\min\{2,\delta-\lambda\}(\beta+1)
\end{equation}
instead of \Eqref{eq:mainerrorestimateexp}.  Given \Eqref{eq:mainerrorestimateexp2} as before, we conclude from this
that if \(0<\delta-\lambda< 2\), and therefore
\(\sigma_{num}=(\delta-\lambda)(\beta+1)\), it follows that
\(\sigma_{num}<\sigma_{cont}\). In the
same way, if \(\delta-\lambda\ge 2\), and therefore
\(\sigma_{num}=2\beta+2\), it also follows that
\(\sigma_{num}<\sigma_{cont}\). 
In conclusion, the numerical approximation is
  therefore {\sl never asymptotically balanced if $\beta<0$}, while the total
  approximation error is {\sl always dominated by the
  numerical error asymptotically.} 
  Using the same estimate for the numerical work
\eqref{eq:estimateruntime}, we find that \textsl{the asymptotic efficiency exponent} is
therefore $\min\{\delta-\lambda,2\}$ \textsl{independently of $\beta$ if $\beta<0$}. 
The numerical efficiency which can be achieved for $\beta<0$ is therefore never better than the one which we can achieve for $\beta\ge 0$.

\begin{figure}[t!]

  \hspace*{\fill}%
  \begin{minipage}[t]{.49\textwidth}
    \centering
    \vspace{0pt}
    \includegraphics[height=5.3cm]%
    {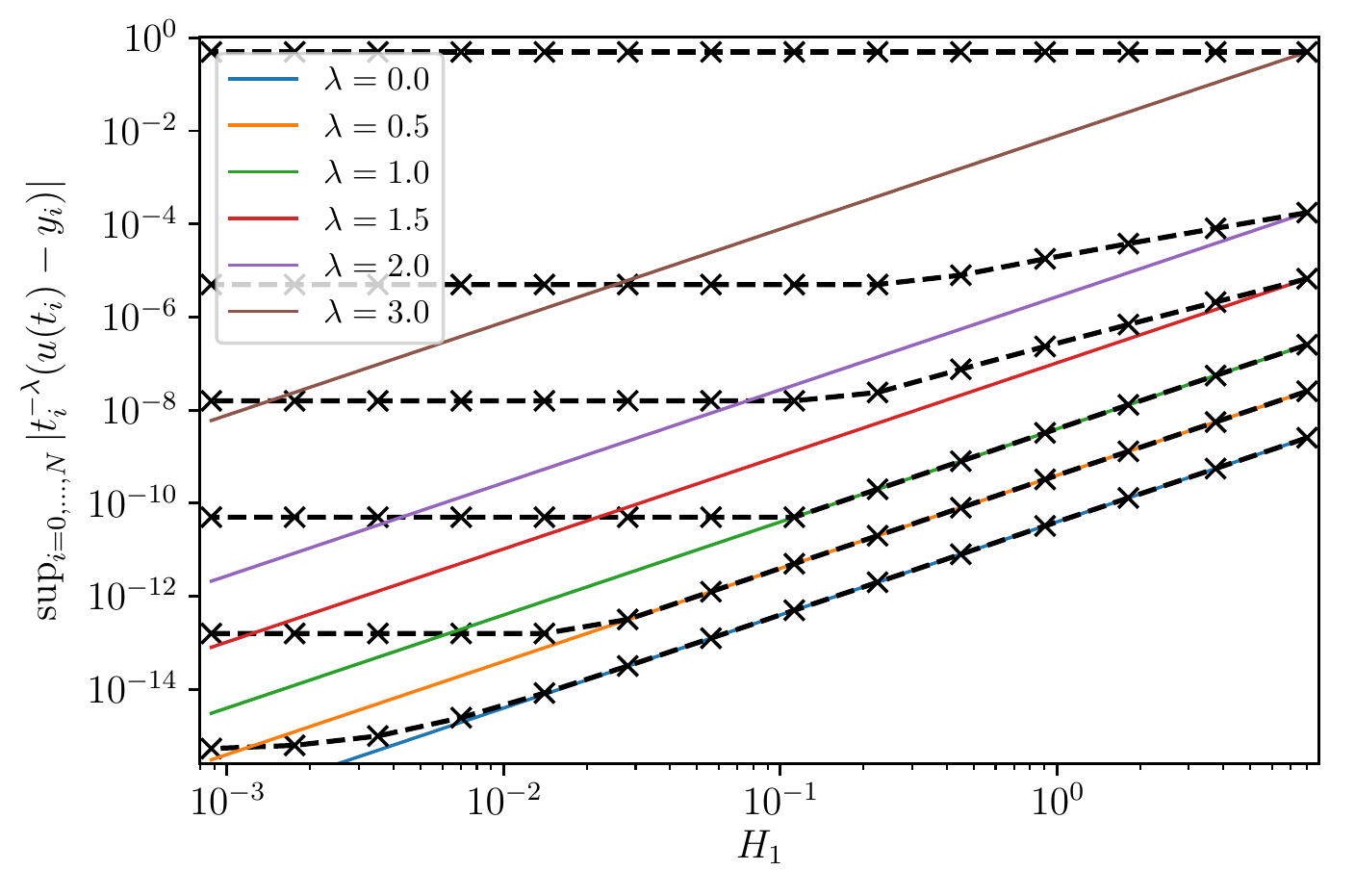}      
  \end{minipage}%
  \hfill
  \begin{minipage}[t]{.49\textwidth}
    \centering   
    \vspace{0pt}
    \includegraphics[height=5.3cm]%
    {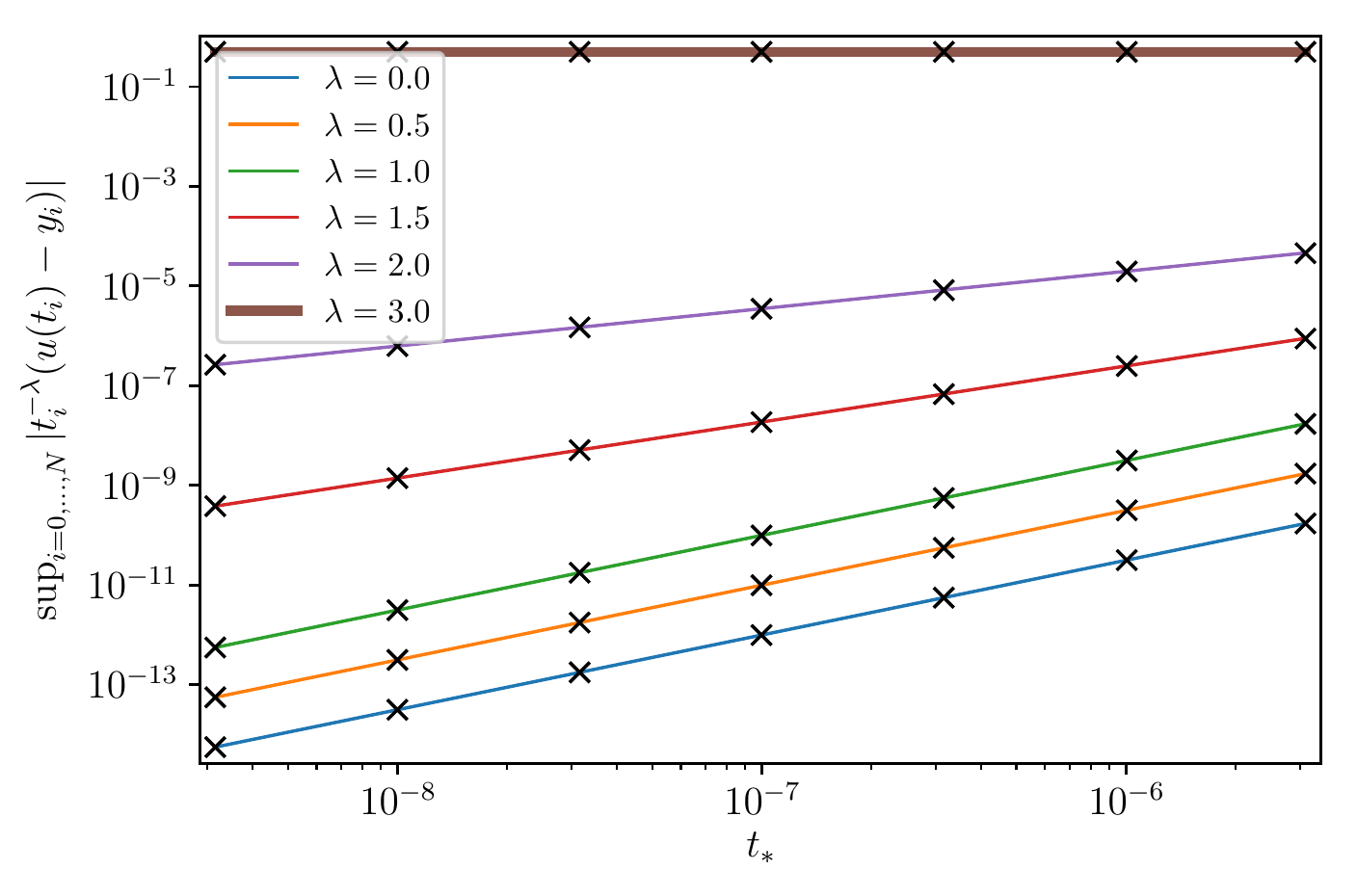}     
  \end{minipage}%
  \hspace*{\fill}

\caption{The same as \Figref{fig:numN1} but with  $\beta=-0.25$. We choose the fixed values
      $t_*=10^{-5}$ for the first plots and $H_1=5$ for the second plot.}
  \label{fig:numN10}
\end{figure} 

\begin{figure}[t]

  \hspace*{\fill}%
  \begin{minipage}[t]{.49\textwidth}
    \centering
    \vspace{0pt}
    \includegraphics[height=5.3cm]%
    {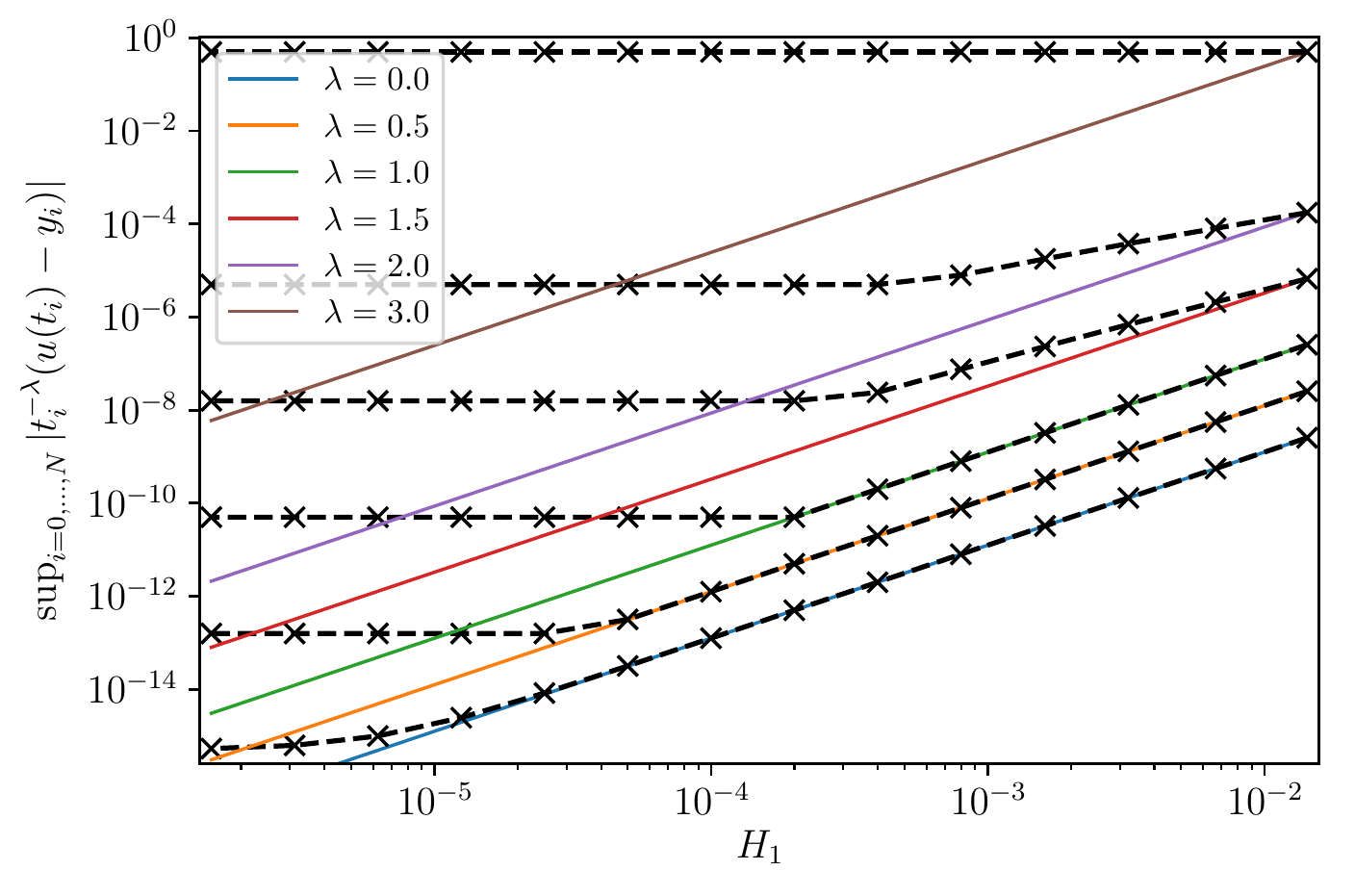}      
  \end{minipage}%
  \hfill
  \begin{minipage}[t]{.49\textwidth}
    \centering   
    \vspace{0pt}
    \includegraphics[height=5.3cm]%
    {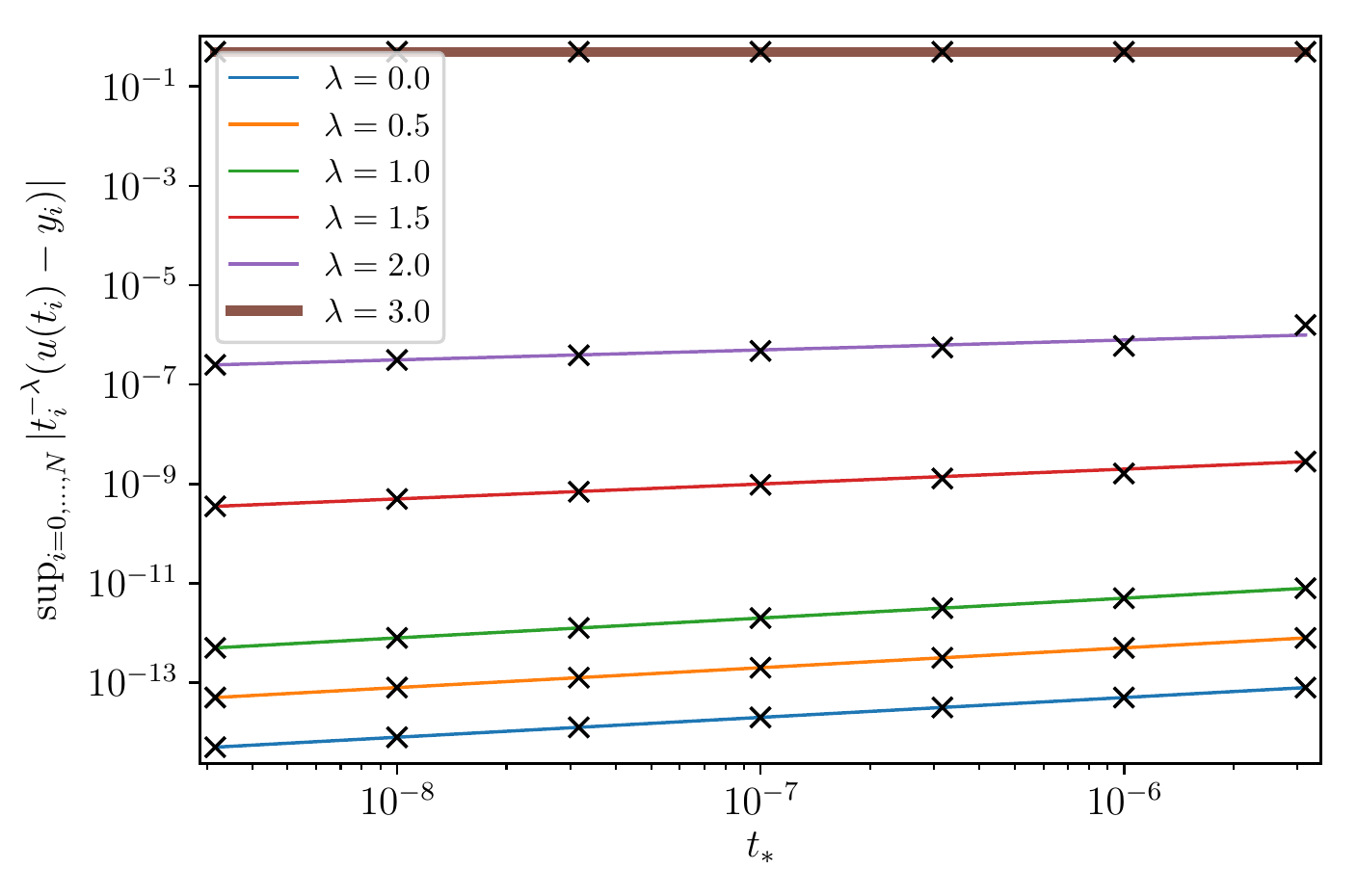}     
  \end{minipage}%
  \hspace*{\fill}

  \caption{The same as \Figref{fig:numN1} but with  $\beta=-0.8$. We choose the fixed values
      $t_*=10^{-5}$ for the first plots and $H_1=10^{-4}$ for the second plot.}  
  \label{fig:numN12}
\end{figure}

Let us confirm these findings numerically now by first picking up the values 
\(p=1.5\), $\eta=0$, and \(\beta=-0.25\); the results
are presented in \Figref{fig:numN10}. The results in
\Figref{fig:numN12} show the same numerical experiments, but with \(\beta=-0.8\). While everything is in agreement, we clearly notice how much the slower convergence rates are here than in the case $\beta=-0.25$ or even the case $\beta=0$ ---as expected.


\section{Application to fluid flows on a Kasner background}
\label{sec:mainnumapplications}
 
\subsection{Numerical algorithm}

\label{sec:mainnumericalsetup}

We are finally in a position to apply the theory developed \Sectionref{sec:errorestimates} to the Euler
equations \eqref{eq:Eulereqssymmhyp1}--\eqref{eq:Eulereqssymmhyp1G} in the  sub-critical case $\Gamma>0$. We proceed by considering a \emph{semi-discretized approximation} of the Euler equations on a Kasner background, formulated as follows. 

We are given $n$ equidistant
spatial grid points $x_0,\ldots x_{n-1}$ on $\mathbb T^1$ with
$x_k=\frac{2\pi}{n} k$ for $k=0,\ldots,n-1$. All $x$-dependent quantities, in particular the unknowns of the Euler equation, are then expanded in a finite Fourier series at each time $t$, so that the
number of real coefficients for each function agrees with the number
of grid points. Each $x$-dependent function is then represented at each time equivalently by the $n$-dimensional vector of its grid point values as well as the $n$-dimensional vector of Fourier coefficients. In such a \emph{pseudo-spectral method,} we almost exclusively work with the former representation with the only exception of spatial derivatives which are approximated using the latter. As a consequence, the Euler equations \eqref{eq:Eulereqssymmhyp1}--\eqref{eq:Eulereqssymmhyp1G} turn into a  {\sl system
of $2n$ ordinary differential equations in time} which are coupled together and take the form
\begin{equation}
  \label{eq:semidiscretisedEuler}
  t\partial_t\Vb(t)-A\Vb(t)=f(t,\Vb(t)). 
\end{equation}
Here, the unknown $\Vb$ has $2n$ components defined by the values at each grid point of the two unknown functions $V^0$ and $V^1$, and the $2n\times 2n$-matrix $A$ reads 
\[A=\mathrm{diag}\,\Bigl(
    \underbrace{\Gamma,\ldots,\Gamma}_{\text{$n$ times}},
    \underbrace{2\Gamma,\ldots,2\Gamma}_{\text{$n$ times}}\Bigr),\]
and where $f$ is some (rather involved) $2n$-dimensional expression which can be derived from
the Euler equations \eqref{eq:Eulereqssymmhyp1}--\eqref{eq:Eulereqssymmhyp1G}. Observe that while $A$ is diagonal, it is a consequence of the 
pseudo-spectral nature of the spatial discretization that $f$ is significantly {\sl non-sparse.}
 More details about such (pseudo)-spectral \emph{method of lines} (MoL) techniques can be found in, for instance, \cite{boyd1989}. 
 
 At this stage we have not (yet) discretized the time variable, and \Eqref{eq:semidiscretisedEuler} is therefore referred to as a semi-discrete approximation of the Euler equations. We have then reduced the evolution problem for the Euler equations to the type of Fuchsian system we have studied in the previous sections of this paper. 

We are interested in numerical approximations for solving the singular initial value problem for \Eqref{eq:semidiscretisedEuler}. To this end, given smooth asymptotic data $V^0_*>0$ and $V^1_{*}$, it takes the form
\Eqref{eq:FuchsianODEgen}, i.e.,
\begin{equation}
  \label{eq:semidiscreteEulerabstr}
  t\partial_t\ub(t)-A\ub(t)=\fb(t,\ub(t))=f(t, \VbLOT(t)+\ub(t))-t\partial_t \VbLOT(t)-A \VbLOT(t),
\end{equation}
 for the remainder and leading-order term, respectively,
\be
\aligned 
   \ub(t)&=\Vb(t)-\VbLOT(t),\\ 
   \VbLOT(t)&=\Bigl(
    V^0_*(x_0)t^\Gamma,\ldots,^0_*(x_{n-1})t^\Gamma,
    V^1_*(x_0)t^{2\Gamma},\ldots, V^1_*(x_{n-1})t^{2\Gamma}\Bigr).
\endaligned
\ee

Our strategy is as follows: 
  \begin{enumerate}
  \item Use the methods in \Sectionref{sec:errorestimates} to calculate accurate
    numerical approximations for the semi-discretized Euler equations with the
    asymptotics \eqref{eq:fluidspecialasympt3}. In other words, numerically
    solve the singular initial value problem for some given asymptotic data $V_*$ and $V_{**}$.
    
  \item Having constructed a numerical approximation $\yb(t)$ of the solution $\ub(t)$ (to the singular initial value problem) up to some time $t=T>0$, we then set
    \begin{equation}
      \label{eq:perturbedID}
      \VbCD=\VbLOT(T)+\yb(T)+\epsilon \Gb
    \end{equation}
for a \emph{perturbation parameter} $\epsilon\in\mathbb R$ and for some fixed $\Gb$ representing the grid values of some smooth \emph{perturbation function}. 

\item We then calculate a numerical approximation $\Vb_\epsilon$ of the solution backwards in time towards $t=0$ for the Cauchy problem associated with \Eqref{eq:semidiscretisedEuler} when the Cauchy data $\VbCD$ are given by \Eqref{eq:perturbedID} and are imposed at $t=T$.

\item Finally we numerically compare the limits of the first $n$ components of $t^{-\Gamma}\Vb_\epsilon$ and the last $n$ components of $t^{-2\Gamma}\Vb_\epsilon$ at $t=0$ with $\VbLOT$ and study how the difference depends on the size of $\epsilon$.
\end{enumerate}
Recall that \cite{beyer2019a} only addresses the continuous analogue of the limit of the first $n$ components of $t^{-\Gamma}\Vb_\epsilon$, but not of the last $n$ components of $t^{-2\Gamma}\Vb_\epsilon$. We shall therefore particularly emphasize these last $n$-components below.
We emphasize that our (Python based) implementation of the previous algorithm has passed several basic tests before we proceeded with the following analysis.


\subsection{Numerical investigations}
\label{sec:mainnumericalresults}

We choose the following parameters for
the Kasner background spacetime and the Euler fluid
\begin{equation}\label{eq:Kasnerbackgroundparam}
\gamma=\frac{5}3, \qquad K=\frac12.
\end{equation}
We therefore have $\Gamma\approx0.73$. We pick up the number of grid points to be $n=80$.

\paragraph{Numerical Fuchsian analysis of the singular initial value problem.}

  The first step is now to construct solutions $\ub$ of the singular
initial value problem of the semi-discrete Euler equations \Eqref{eq:semidiscreteEulerabstr} on the above Kasner
background. All numerical results here are shown for 
\begin{equation}\label{eq:mainasymptoticdata}
    V^0_*(x)=1,\qquad V_{*}^1(x)=\frac 32\cos x.
\end{equation}
As the numerical time integrator we choose RK2 as before. As suggested by the results in \Sectionref{sec:discparam}, we 
restrict to $\beta=\eta=0$ here and choose
$\alpha_i=\frac{H_1 t_*}{t_i}$, for all $i=0,\ldots,n-1$,
in agreement with \Eqsref{eq:alpharestr} and \eqref{eq:boundednessalpha}.

\begin{itemize}

\item 
In contrast to \Sectionref{sec:Numtestproblem}, clearly we {\sl do not have an exact solution} for the singular initial value problem at our disposal. Total approximation errors are therefore estimated with respect to \emph{numerically obtained reference solutions}. In our sequence of numerical solutions we shall always choose the highest resolution solution as that reference solution. 

\item Note also that we shall always multiply all Euclidean vector norms $|\cdot|$ in this section by the factor $\sqrt{2\pi/n}$. This guarantees that these norms approximate the spatial $L^2$-norm in the limit $n\to \infty$ correctly. 
 
\end{itemize}

\begin{figure}[t!]
    \begin{center}
        \includegraphics[height=5.3cm]{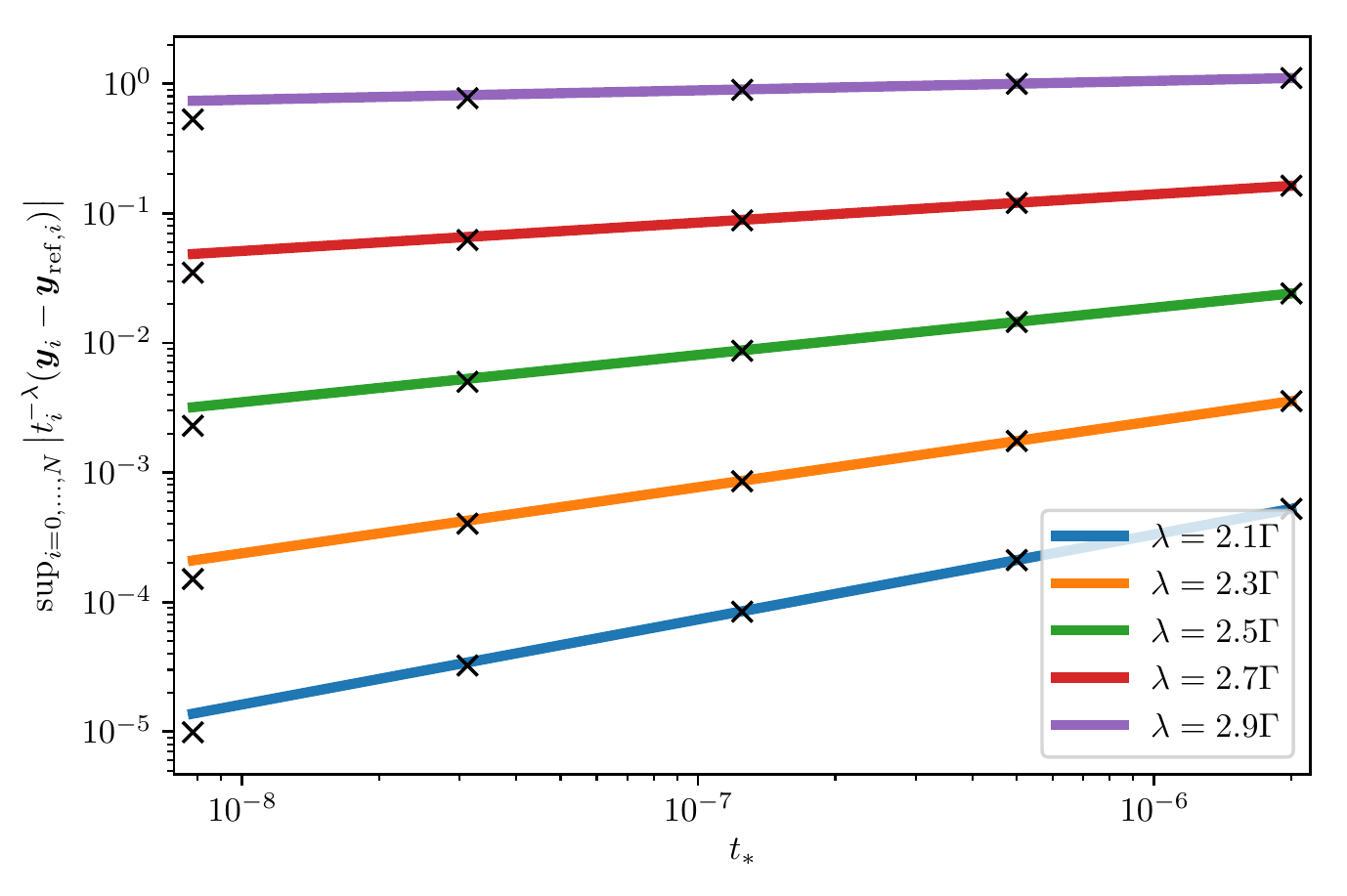}
    \end{center}
    \caption{Solutions of the singular initial value problem of the Euler equations on the Kasner background given by \Eqsref{eq:Kasnerbackgroundparam} and \eqref{eq:mainasymptoticdata}: Total approximation errors obtained for $n=80$, $H_1=3$, $T=3.98\cdot 10^{-4}$ and various values of $\lambda$. The total approximation errors are marked by $\times$ while the continuous reference curves show the theoretical
      $O(t_*^\sigma)$ behavior with $\sigma=\delta-\lambda$.}  
    \label{fig: Euler_53_0}
\end{figure}

It follows from our theoretical analysis in \cite{beyer-PLF-2017} that $\delta=3\Gamma$ and $\mu_0\in (2\Gamma, 3\Gamma)$. 
  Given $\Gamma\approx 0.73$ and $\beta=\eta=0$, \Theoremref{theorem:RK2} implies
that $\sigma_{num}=\min\{2,\delta-\lambda\}$. Since $\lambda$ is
required to be between $2\Gamma$ and $\delta$, it follows that
$\sigma_{num}=\delta-\lambda=\sigma_{cont}=\sigma$ from \Eqref{eq:mainerrorestimateexp2}. The numerical and
the continuum approximation errors are therefore always asymptotically
balanced. In \Figref{fig: Euler_53_0} we confirm this numerically with
$H_1=3$, $T=3.98\cdot 10^{-4}$ and various values of $\lambda$. Our choice $H_1=3$ is close to the ``optimal value'' found using the algorithm suggested  at the end of \Sectionref{sec:discparam}.

\begin{figure}[t!]

  \hspace*{\fill}%
  \begin{minipage}[t]{.49\textwidth}
    \centering
    \vspace{0pt}
    \includegraphics[height=5.3cm]%
    {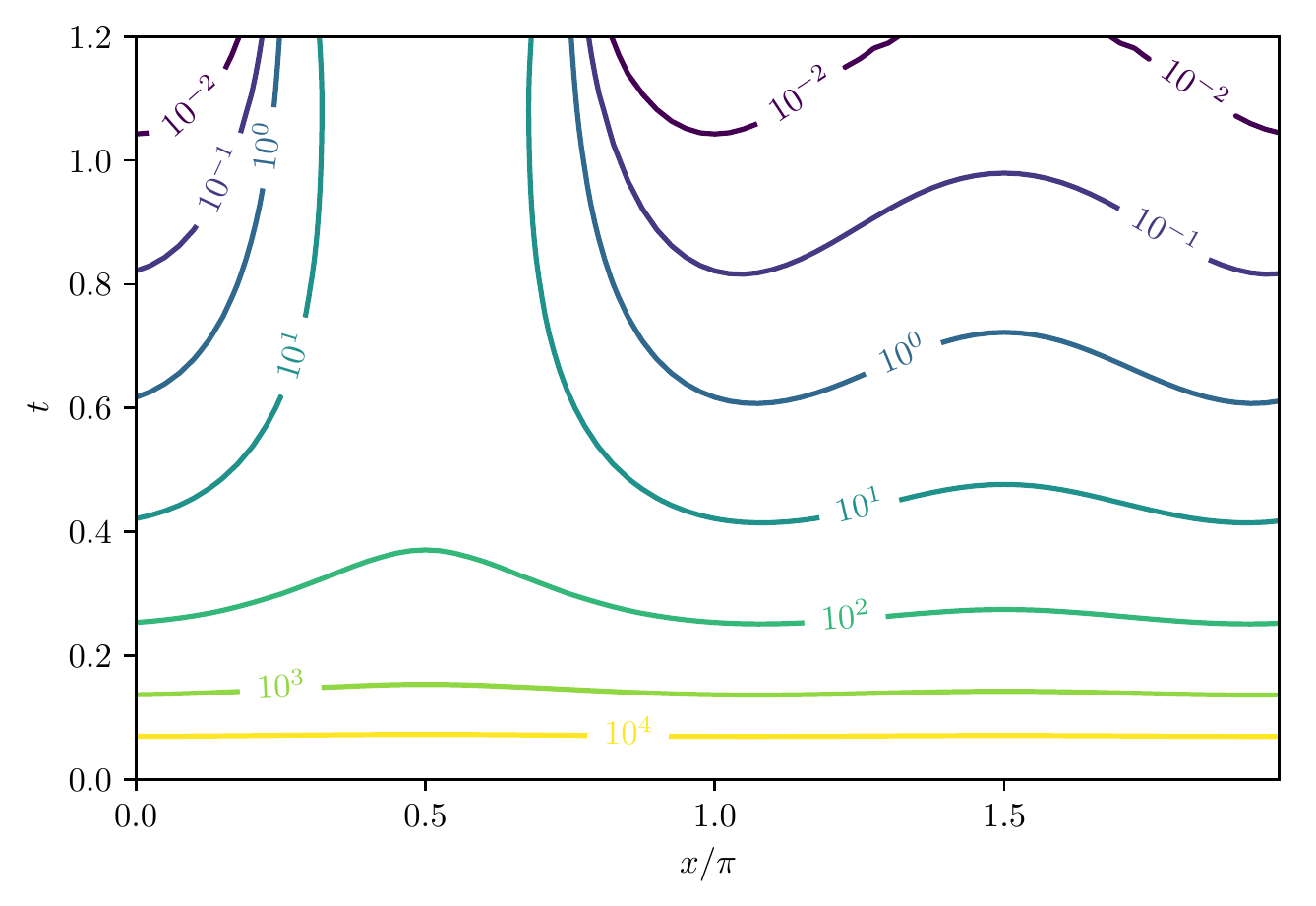}      
  \end{minipage}%
  \hfill
  \begin{minipage}[t]{.49\textwidth}
    \centering   
    \vspace{0pt}
    \includegraphics[height=5.3cm]%
    {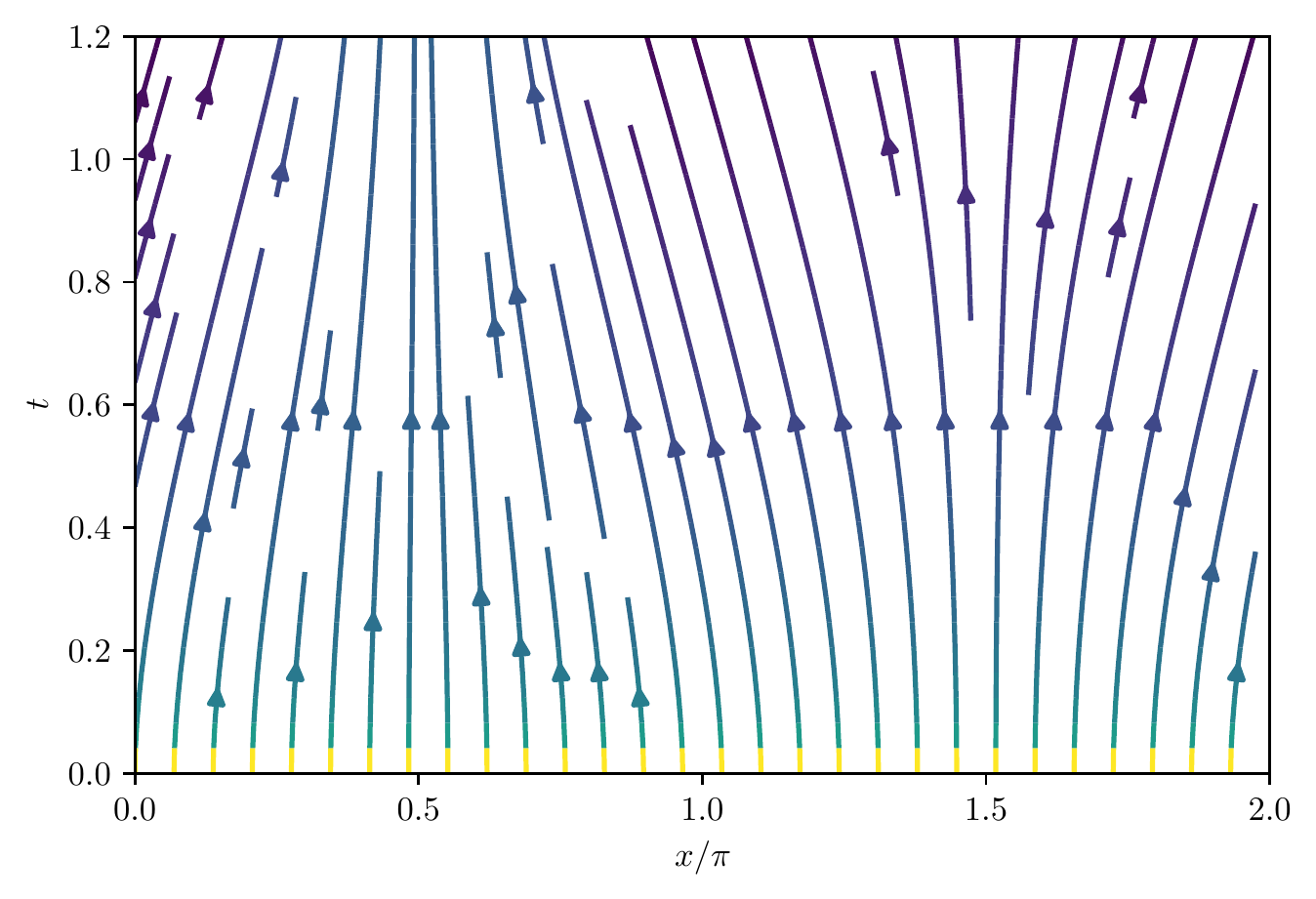}     
  \end{minipage}%
  \hspace*{\fill}
\caption{Solutions of the singular initial value problem of the Euler equations on the Kasner background given by \Eqsref{eq:Kasnerbackgroundparam} and \eqref{eq:mainasymptoticdata}: Contour plot of the fluid density (left plot) and the flow lines (right plot) for the  numerical approximation of the solution of the singular initial value problem given by the smallest $t_*$-value in \Figref{fig: Euler_53_0}.}  
  \label{fig:Euler_71_1}
\end{figure} 

Before we continue to investigate perturbations of this solution of the singular initial value problem  (and provide an answer to the main question of interest), let us first present two plots regarding the physical properties of the resulting fluid in \Figref{fig:Euler_71_1}. The left-hand plot shows the level sets of the function $\rho$ given by \Eqref{eq:physicsquantitiesfluid} on our Kasner spacetime. This plot confirms that $\rho$ is unbounded in the limit $t\searrow 0$. Close to $t=1.2$, the numerical solution becomes very noisy and cannot be trusted any further as one can see from the slightly ``wriggly'' $10^{-2}$ contour line. We have not yet investigated whether this is an actual physical phenomenon, which our pseudo-spectral code is not able to resolve, or whether it is a purely numerical issue. The right-hand plot in \Figref{fig:Euler_71_1} shows the flow lines of the fluid solution where the normalized vector field $U$ is given in \Eqref{eq:physicsquantitiesfluid}. It is clear from these plots that this solution to the singular initial value problem is {\sl highly non-trivial and very inhomogeneous,} so it provides us with a good test case.


\paragraph{Numerical perturbations of the solution of the singular initial value problem.}
Having obtained sufficiently accurate numerical approximations of solutions of the singular initial value problem, the next steps are now (see steps 2, 3 and 4 in \Sectionref{sec:mainnumericalsetup}) to calculate 
{\sl perturbations} by evolving perturbed Cauchy data in \Eqref{eq:perturbedID} determined by a perturbation parameter denoted by $\epsilon\in\mathbb R$ and a fixed $2n$-dimensional vector $\Gb$ backwards from the initial time $t=T$ towards $t=0$. As before we choose $T=3.98\cdot 10^{-4}$. For the purpose of this presentation, we pick the $2n$-dimensional vector $\Gb$ from the grid point values of the function
\begin{equation}
G(x)=(0,\cos x).
\end{equation} 
With this choice of function we can assure that the fluid vector is initially timelike for all values of $\epsilon$ sufficiently close to zero.

In order to perform the numerical calculations backwards in time from the starting time $t=T=3.98\cdot 10^{-4}$ towards the singularity $t=0$, we use the Livermore Solver for Ordinary Differential
Equations (LSODE) in \cite{hindmarsh1980} with an absolute error tolerance of $10^{-11}$ instead of the RK2-time integrator (but otherwise we use the same numerical infrastructure). We can justify this use by pointing out that this is now nothing but a regular Cauchy problem.
With this error tolerance we can guarantee that the numerical errors for the backward evolutions are negligible in comparison to all of the other errors, and we can therefore treat these like ``exact solutions'' $\Vb_\epsilon$ for our purpose.

We present our results for a range of values for $\epsilon$ from $0$ to $0.003$. As discussed before, we know from the results in \cite{beyer2019a} that the solutions of the Euler equations with perturbed data must extend all the way to $t=0$ if the perturbation parameter is sufficiently close to zero. Our numerics suggest that this is no longer true if $\epsilon$ exceeds the value $0.003$. 

\begin{figure}[t!]
  \hfill
  \begin{minipage}[t]{.49\textwidth}
    \centering
    \vspace{0pt}
    \includegraphics[height=5.3cm]%
    {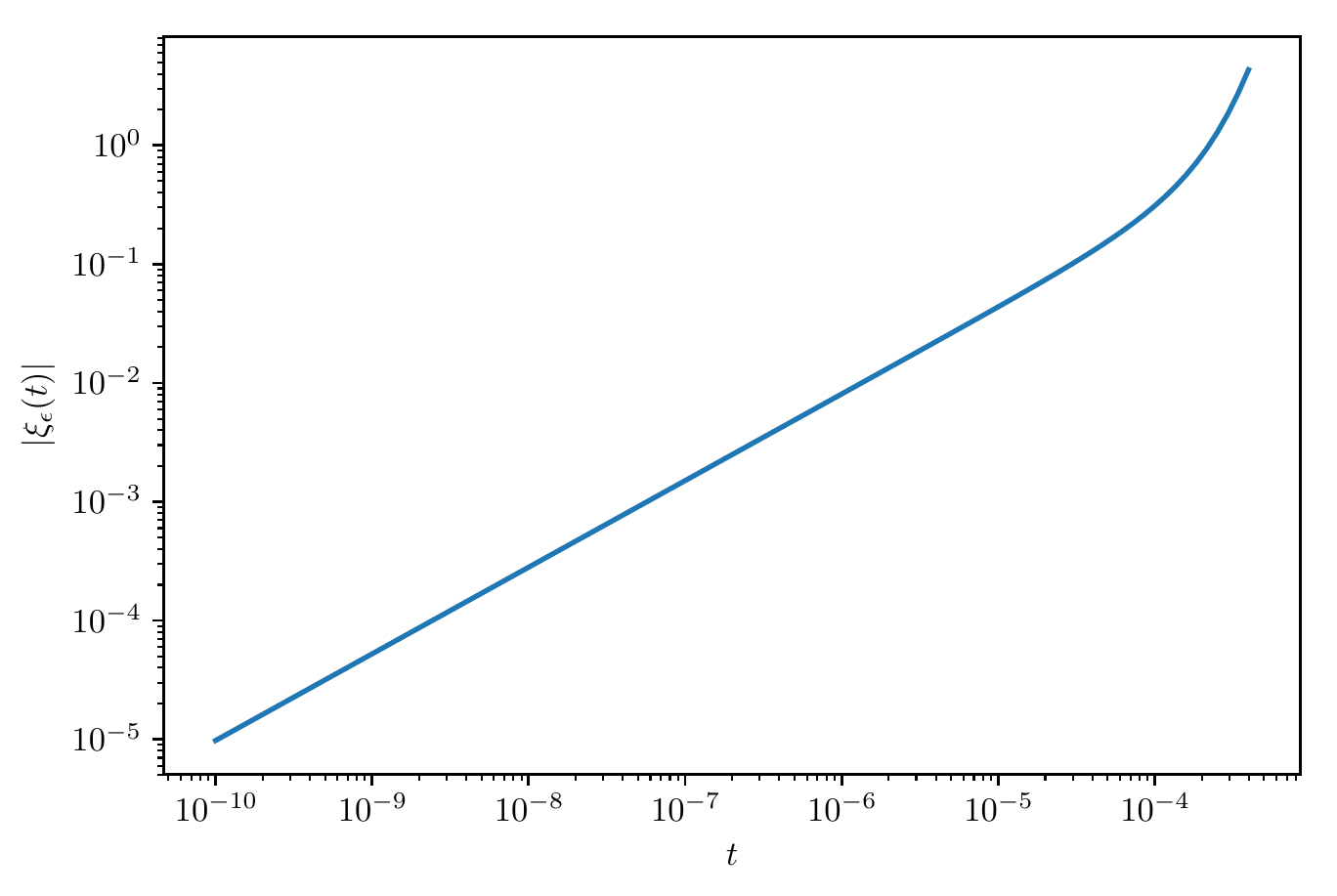}           
  \end{minipage}%
  \hspace*{\fill}

  \begin{minipage}[t]{.49\textwidth}
    \centering   
    \includegraphics[height=5.3cm]%
    {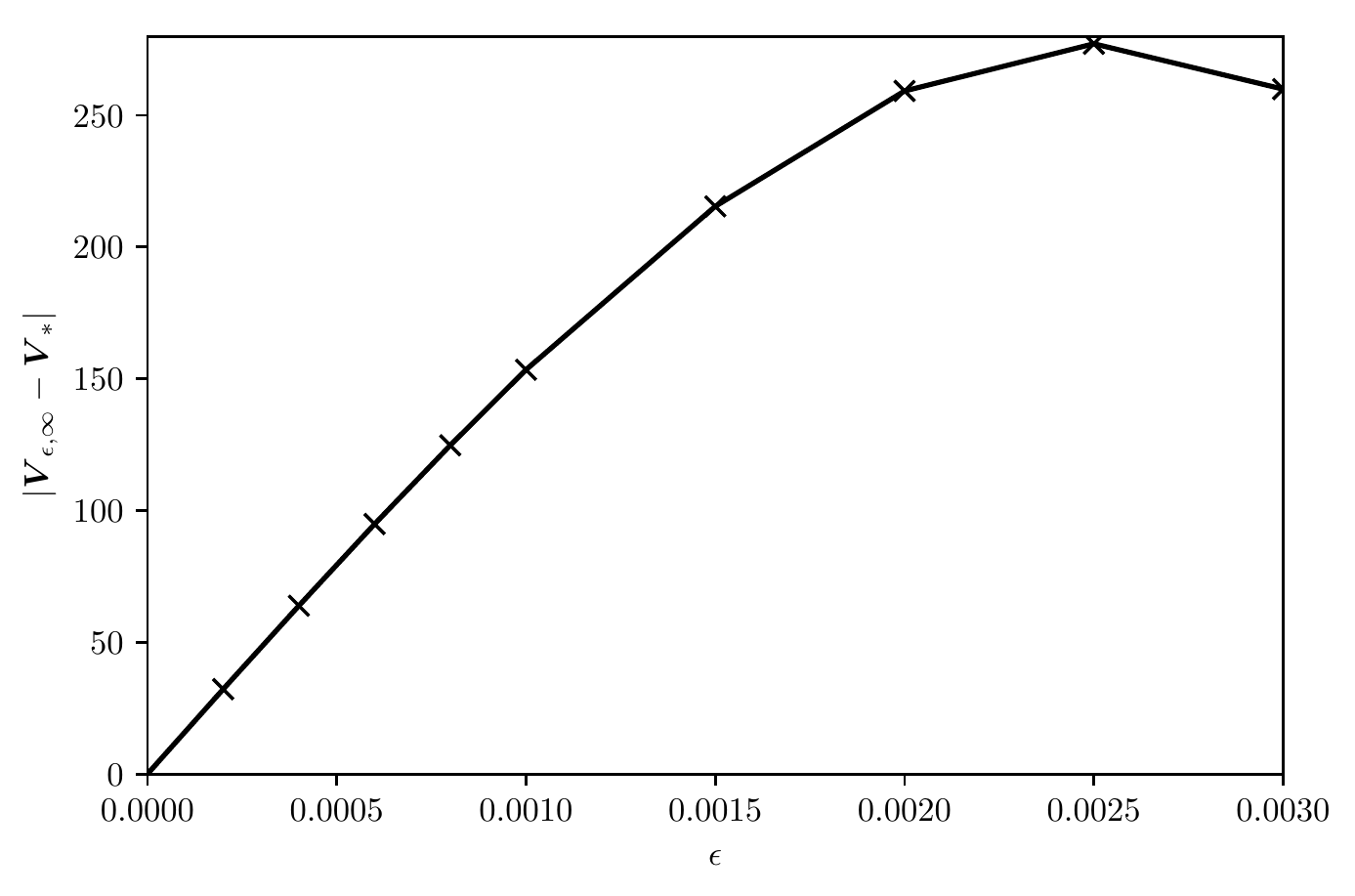}         
    \label{fig:numN2}
  \end{minipage}%
  \begin{minipage}[t]{.49\textwidth}
    \centering
    \includegraphics[height=5.3cm]%
    {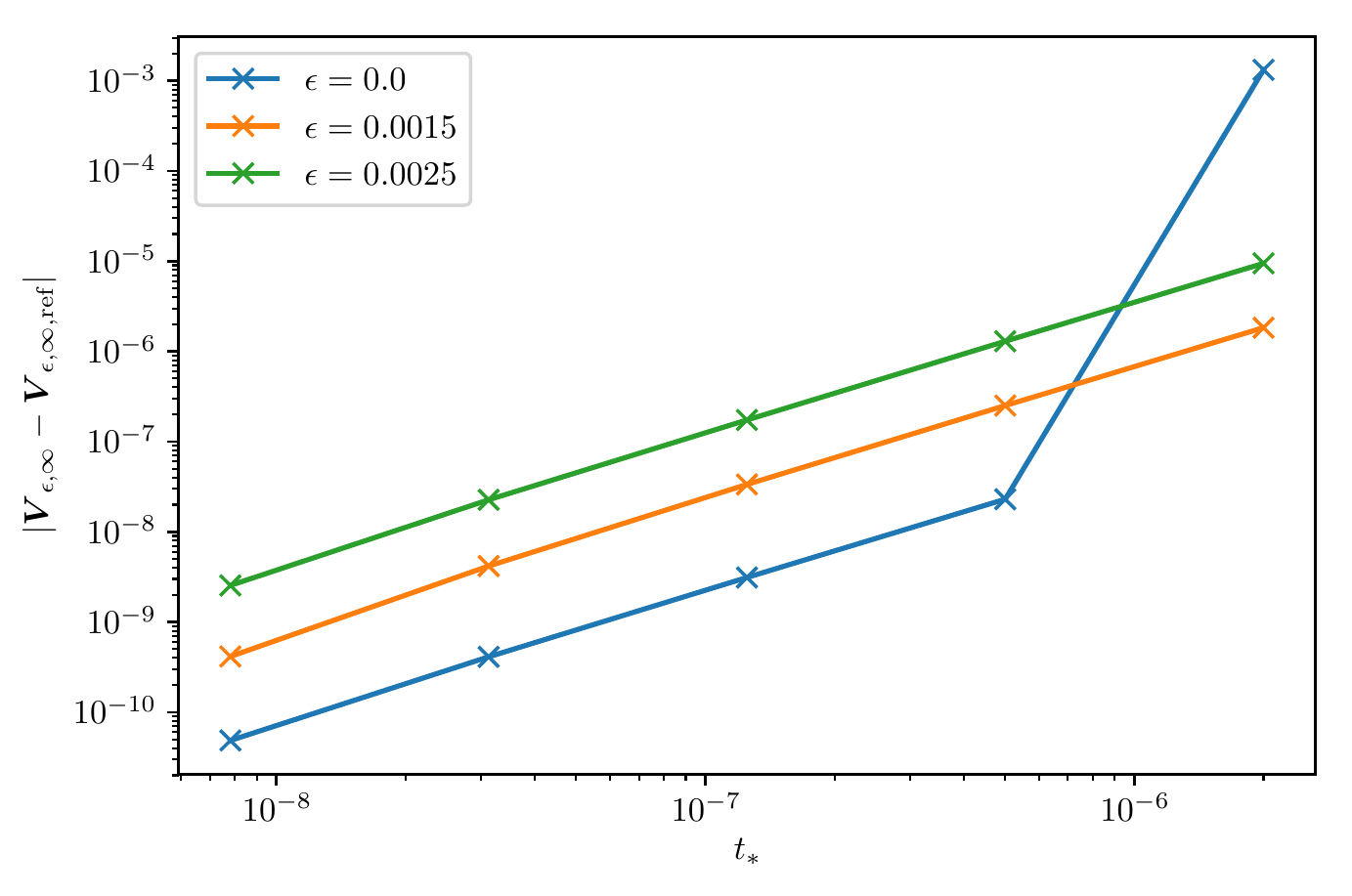}  
  \end{minipage}

\caption{Perturbations of the solutions singular initial value problem of the Euler equations on the Kasner background given by \Eqsref{eq:Kasnerbackgroundparam} and \eqref{eq:mainasymptoticdata}: The first plot shows the function $\boldsymbol \xi_{\epsilon}(t)$ defined in \Eqref{eq:xiepsdef} for the largest $\epsilon$ and the smallest $t_*$. The second plot compares the limit $\boldsymbol V_{\epsilon,\infty}$ defined in \Eqref{eq:limitEuler} to the asymptotic data as a function of $\epsilon$ for the smallest value of $t_*$. The third plot confirms that the numerical approximation used to determine $\boldsymbol V_{\epsilon,\infty}$ converges in the limit $t_*\to  0$.}   
\label{fig:Euler_67_0}
\end{figure}

We proceed now by demonstrating that the numerical solutions  $\Vb_\epsilon$ suggest that 
\begin{equation}
\label{eq:limitEuler}
\mathrm{diag}\left(t^{-\Gamma},\ldots, t^{-\Gamma}, t^{-2\Gamma},\ldots, t^{-2\Gamma}\right)\cdot\boldsymbol V_{\epsilon}(t)=\boldsymbol V_{\epsilon,\infty}+\boldsymbol U_{\epsilon}(t),
\end{equation}
close to $t=0$
where $\boldsymbol V_{\epsilon,\infty}$ is a $2n$-vector 
(which depends on $\epsilon$ but not on $t$)
and where $\boldsymbol U_{\epsilon}(t)$ is a $2n$-vector-valued function $\boldsymbol U_{\epsilon}(t)$ which decays like a positive power of $t$  in the limit $t\searrow 0$. In order to support this claim we 
show that
\begin{equation}
  \label{eq:xiepsdef}
\boldsymbol \xi_{\epsilon}(t)=t\partial_t\left(\mathrm{diag}\left(t^{-\Gamma},\ldots, t^{-\Gamma}, t^{-2\Gamma},\ldots, t^{-2\Gamma}\right)\cdot\boldsymbol V_{\epsilon}(t)\right)
\end{equation}
behaves like a positive power of $t$  in the limit $t\searrow 0$; see the first plot in \Figref{fig:Euler_67_0}, where we  numerically calculate this function for the perturbation of the solution given by the smallest value of $t_*$ above (i.e., our best approximation of the solution of the singular initial value problem) and the largest value of $\epsilon=0.003$ (i.e., our largest considered perturbation). Observe carefully that this is a {\sl double-log plot.}

 If this is the case, the claim in \Eqref{eq:limitEuler} must follow, and it remains to estimate $\boldsymbol V_{\epsilon,\infty}$ and then to compare it to the asymptotic data $V_*$ and $V_{**}$. We estimate $\boldsymbol V_{\epsilon,\infty}$ by choosing a sufficiently small \emph{read off time} $t_{RO}$ and then setting
\be
\boldsymbol V_{\epsilon,\infty}\approx \left.\mathrm{diag}\left(t^{-\Gamma},\ldots, t^{-\Gamma}, t^{-2\Gamma},\ldots, t^{-2\Gamma}\right)\cdot\boldsymbol V_{\epsilon}(t)\right|_{t=t_{RO}}.
\ee
For this presentation we have chosen $t_{RO}=10^{-10}$. The results are presented in the second and third plots of \Figref{fig:Euler_67_0}.
The main observation is that the second plot confirms that $\boldsymbol V_{\epsilon,\infty}$ depends continuously on $\epsilon$, i.e., it approaches the asymptotic data in the limit $\epsilon\to  0$. A second interesting observation is the local maximum at $\epsilon\approx 0.0025$. Given that the absolute values shown in this second plot are also quite large,  this indicates that our numerical results are far from any regime in which ``linear'' perturbation theory would be applicable despite the apparent smallness of the chosen value for the perturbation parameter $\epsilon$. 
Finally, the third plot of \Figref{fig:Euler_67_0} is just a confirmation that  our numerical approximation of the limits $\boldsymbol V_{\epsilon,\infty}$ converge as expected as $t_*$ goes to zero.

\small 

\paragraph*{Acknowledgments} 
The second author (PLF) was partially supported by the Innovative Training Network (ITN) grant 642768 (ModCompShock), and by the Centre National de la Recherche Scientifique (CNRS).

\end{document}